\documentclass[aps,prb,preprint,showpacs,amsmath,superscriptaddress]{revtex4}
\usepackage{graphicx}
\usepackage{color}
\usepackage{amsmath}
\usepackage{esint}
\usepackage[american]{babel}
\usepackage{ulem}

\begin{document}

\newcommand{\LK}[1]{\textcolor{blue}{#1}}

\def\K{{\bf{K}}}
\def\Q{{\bf{Q}}}
\def\Gbar{\bar{G}}
\def\tk{\tilde{\bf{k}}}
\def\k{{\bf{k}}}
\def\Kp{{\bf{K}}^{\prime}}
\def\tp{t^{\prime}}
\def\Jeff{J_{\textrm{\textit{eff}}}}
\def\half{\frac{1}{2}}
\def\sixth{\frac{1}{6}}
\def\third{\frac{1}{3}}
\def\wn{\omega_n}
\def\wnp{\omega_n'}
\def\Jp{J^{\prime}}
\def\Up{U^{\prime}}
\def\ba{\begin{align}}
\def\ea{\end{align}}
\def\Ham{\mathcal{H}}
\def\Emax{E$_\textrm{max}$}
\def\Amax{A$_\textrm{max}$}
\newcommand{\red}{\textcolor[rgb]{1.00,0.00,0.00}}
\newcommand{\blue}{\textcolor[rgb]{0.00,0.00,1.00}}

\title{Theoretical description of high-order harmonic generation in solids}

\author{A. F. Kemper}
\email[]{kemper@stanford.edu}
\affiliation{Stanford Institute for Materials and Energy Science, SLAC National Accelerator Laboratory, Menlo Park, CA 94025, USA}
\affiliation{Geballe Laboratory for Advanced Materials, Stanford University, Stanford, CA 94305, USA}

\author{B. Moritz}
\affiliation{Stanford Institute for Materials and Energy Science, SLAC National Accelerator Laboratory, Menlo Park, CA 94025, USA}
\affiliation{Department of Physics and Astrophysics, University of North Dakota, Grand Forks, ND 58202, USA}

\author{J.~K. Freericks}
\affiliation{Department of Physics, Georgetown University, Washington, DC 20057, USA}

\author{T. P. Devereaux}
\affiliation{Stanford Institute for Materials and Energy Science, SLAC National Accelerator Laboratory, Menlo Park, CA 94025, USA}
\affiliation{Geballe Laboratory for Advanced Materials, Stanford University, Stanford, CA 94305, USA}

\date{\today}

\begin{abstract}
We consider several aspects of high-order harmonic generation
in solids: the effects of elastic and inelastic scattering; varying pulse characteristics; and inclusion of material-specific parameters through a realistic band structure. We reproduce many observed characteristics of
high harmonic generation experiments in solids including the formation of only odd harmonics in inversion-symmetric materials, and the nonlinear formation of high harmonics with increasing field.  
We find that the harmonic spectra are fairly robust against elastic and inelastic scattering.
Furthermore, we
find that the pulse characteristics play an important role in determining the harmonic spectra.
\end{abstract}

\maketitle

\section{Introduction}
The rapid development of time-resolved pump-probe experiments such as angle-resolved photoemission spectroscopy (ARPES)
\cite{a_cavalieri_07,f_schmitt_08,l_perfetti_08}, 
optical reflectivity\cite{d_basov_11} and 
X-ray scattering\cite{a_cavalleri_01,a_rousse_01} has expanded greatly the use of strong ultrashort laser pulses in the study of condensed matter systems. 
The advent of these new experimental techniques requires
theory to properly describe the non-equilibrium physics observed in these measurements.
The strong temporal dependence of these problems makes analysis in the frequency domain difficult which is more common
for problems in equilibrium or within linear-response.
A variety of methods have been developed or adapted to describe the behavior of these systems
directly in the time domain including semiclassical approximations,\cite{t_tritschler_03,d_golde_08,d_golde_11,o_mucke_11} 
Floquet theory\cite{f_faisal_97} and direct
numerical integration of the time-dependent Schr\"{o}dinger equation.\cite{p_devries_90} Each of these methods has limitations varying from lack of numerical feasibility to the inability to explicitly treat time-translation invariance breaking due to the pump and/or probe fields.

The non-equilibrium Keldysh formalism, developed
by Kadanoff and Baym\cite{kadanoff_baym} and Keldysh,\cite{l_keldysh_64}
can treat broken time-translation invariance explicitly making use of the so-called Keldysh contour (see Fig.~\ref{fig:k_contour}). Quantities such as the single-particle Green's function are defined along this contour with multiple arguments that take appropriate real and imaginary times along the contour.
Characteristics of the pump pulse, 
such as the driving frequency or pulse length and shape, as well as material specific tight-binding or more realistic band structure parameters for the equilibrium system can be included directly allowing for a closer match between theoretical calculations and realistic experimental conditions. The \textit{ab initio} inclusion of nonlinear effects of the driving field, rather than treating it as a perturbation, is necessary for a proper description of the electronic system as it is driven out of equilibrium. This Keldysh formalism has been adapted and used to study the effect of electronic correlations on systems out of equilibrium within dynamical mean-field theory,
\cite{v_turkowski_05,j_freericks_06,m_eckstein_08,j_freericks_08,m_eckstein_09,b_moritz_10,m_eckstein_10,b_moritz_11}, continous-time Quantum Monte Carlo\cite{m_eckstein_09b,m_eckstein_10} and cluster perturbation theory\cite{m_balzer_11}. A number of other methods exist to treat correlated and uncorrelated time-domain phenomena, including non-equilibrium exact diagonalization\cite{t_park_86}, numerical renormalization group\cite{f_anders_05} and density matrix renormalization group,\cite{s_white_04,g_alvarez_11} with various benefits and limitations.

An ideal application for this Keldysh approach is in the area of high-order harmonic generation (HHG) where one need only consider the non-linear effects of the pump, and can neglect details associated with probe pulses and the often complicated evaluation of a proper pump-probe measurement cross-sections. High-order harmonics result from non-linear light-matter interactions and in the present example from the manner in which the sample current varies in time due to the pulse characteristics of the driving field. Used in atomic gases for the production of attosecond pulses,\cite{p_antoine_96} HHG opens avenues for time-domain experiments with high temporal resolution, and in particular are short enough to resolve electronic interactions which occur on these fundamental timescales. A second use for high-order harmonics is in the production of a wide spectrum of frequencies not available in the fundamental frequencies of standard lasers.
By applying strong driving fields, the harmonics generated can be harvested
for further use as a pump or probe.

Trains of attosecond pulses that can then be used for further experiments\cite{p_paul_01,m_drescher_01} can be generated through the interaction of a femtosecond laser pulse with a jet of gas. In this process the pulses of light are created in three-steps: (1) field-assisted tunnel ionization of an atom, (2) acceleration of the ionized electron back to the atom by the change in sign of the field as the driving pulse oscillates and (3) creation of a train of high-energy photons as the electron re-collides with the charged ion.\cite{j_krause_92,m_lewenstein_94}
In addition to this generation in gaseous media, HHG can be accomplished through non-linear response in solids.
The availability of ultra-short laser pulses with relatively strong fields has facilitated studies in solid state materials while mitigating the risk of sample damage.
This capability has led to recent observations of HHG in ZnO crystals.\cite{s_ghimire_11} Their bulk crystalline nature makes HHG in these materials fundamentally different from what occurs in atoms or small molecules. The electron ionization in an atom is replaced by electron-hole pair excitation between the conduction and valence bands, respectively. The charged particles are then accelerated in the field with a non-linear response due to the Bloch oscillations which occur when the field strength is large enough, as discussed in some detail by several authors.
\cite{d_golde_11,o_mucke_11}

 This is fundamentally different from the single-atom picture,
where the harmonics are generated by the stimulated ionization and recombination of an electron in the
outer shell of an atom.  In a solid, this would correspond to electron-hole pair excitation and de-excitation.
However, as we will show, high-order harmonics can be generated solely via the acceleration of
the conduction electrons by the field, and the Bragg scattering that occurs repeatedly in high field.\cite{s_ghimire_11}
The end result is HHG of the fundamental frequency of the applied pump pulse. 
Although this has been previously studied for non-interacting systems,
\cite{s_ghimire_11,d_golde_11,o_mucke_11} using the non-equilibrium
Keldysh formalism, we can go beyond the semi-classical approach and consider the effects of elastic
and inelastic scattering on the HHG spectra.

 A better understanding of HHG in solids requires a proper description of the underlying electronic states in the material. In atomic gases, these states have been modeled in a number of ways, from partial-wave expansion of the electron cloud to other descriptions utilizing first-principles methods.\cite{j_krause_92} In solids, Bloch states are a basis in which one can describe the itinerant conduction band electrons and valence band holes.
In the presence of a strong driving electric field,
the states are modified Bloch states that, as the field energy
becomes comparable to the bandwidth, become localized.\cite{g_nenciu_91}.
The behavior covering the full range of fields,
is captured by the inclusion of the field via the Peierls' substitution.

The paper proceeds as follows. In Section II, we present relevant details of the non-equilibrium Keldysh formalism as it pertains to HHG. Section III discusses the effects of impurity and phonon scattering. In section IV we consider the effect of different pulse characteristics on HHG and compare and contrast the resulting HHG spectra for two experimentally relevant cases. Section V focuses on the time structure of the individual harmonics.
Finally, section VI studies the HHG spectra from a more realistic, complex band structure appropriate
for Si.

\section{Method}
\begin{figure}[t!]
	\includegraphics[width=\columnwidth]{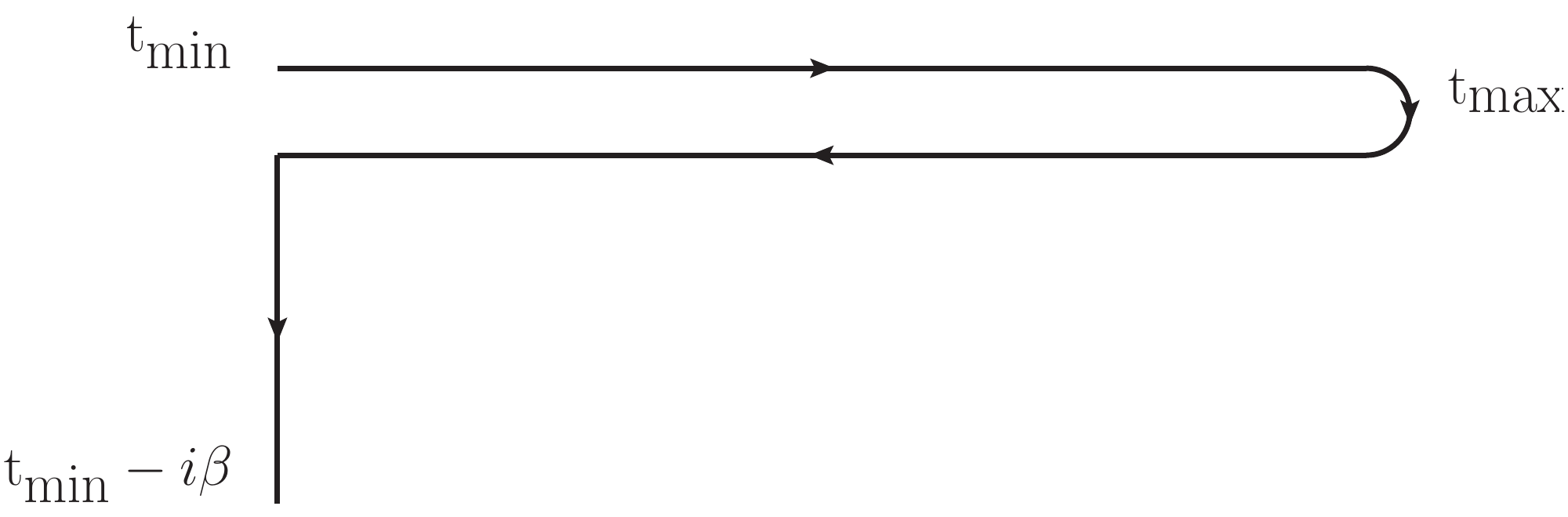}
	\caption{The Keldysh contour used in the calculation of HHG spectra.}
	\label{fig:k_contour}
\end{figure}

We shall consider a system of electrons in the time-domain using the non-equilibrium Keldysh formalism. The appropriate expressions were derived in detail\cite{j_davies_88,v_turkowski_05} previously
and we will present the relevant equations here for the specific case of a spatially uniform field. The system starts in an initial equilibrium state at $t=t_\textrm{min}$, at an initial temperature $T$, and evolves until a final time $t=t_\textrm{max}$.
We describe the system on the Keldysh contour shown in Fig.~\ref{fig:k_contour}. The time arguments of the double-time single-particle Green's functions and related quantities lie on the contour. Properly selecting the time arguments yields the time-ordered, anti-time-ordered, lesser and greater Green's functions,
 and by extension the retarded, advanced, and Keldysh Green's functions (see Ref.~\onlinecite{j_freericks_08}).
For example, the lesser Green's function ($G^<$) is defined with arguments $t$ and $t'$ lying on
the upper and lower branches of the contour, respectively. 
 We use a particular implementation of the non-equilibrium Keldysh formalism where we discretize the Keldysh contour. The discretization is done in steps of $\delta_t$ and
 $i\delta_\tau$ on the real and imaginary parts of the contour, respectively.
This is different from the previous implementation by Turkowski and Freericks, who evaluated the 
temporal integrals for the Dyson equation\cite{g_mahan} directly.\cite{v_turkowski_05} 
However, the direct evaluation of integrals is limited to the case of simple band structures, or, the infinite-dimensional case considered in the dynamical mean-field theory treatment of the Falicov-Kimball model.\cite{l_falicov_69} Here, we numerically evaluate the integrals on a fine grid.

The bare non-equilibrium Green's function $G_0$ under the influence of a driving field can be obtained through
the Peierls' substitution\cite{r_peierls_33} in the standard definition of the double-time Green's function
\begin{align}
G_0(\vec k;t,t') = i \left[ f_{\vec k} - \theta_c(t,t') \right] \exp \left[-i\int_{t'}^t d\bar t\ \epsilon\left(\vec k - \vec A(\bar t)\right)\right] 
\end{align}
where $f_{\vec k}$ is the Fermi function $f_{\vec k}=\left(e^{\epsilon(\vec k)/T}+1\right)^{-1}$, $T$ is the temperature (we work in units where the Boltzmann constant $k_B=1$), $t$ and $t'$ lie on the Keldysh contour, $\theta$ is the contour-ordered version of the standard Heaviside function, $\epsilon(\vec k)$ is the single-particle dispersion, $\vec A(t)$ is the vector potential, which is related to the field by $\vec A(t) = -\int \vec E(t) dt$, where we use $e=c=\hbar=1$ and work in the Hamiltonian gauge. The integral in the exponent runs along the Keldysh contour.

To illustrate the influence of the electronic self-energy on HHG in solids, we include the effect of impurities and electron-phonon scattering.
For impurities, we consider dilute point-like scatterers, for which the self-energy is\cite{a_jauho_84}
\begin{align}
\Sigma(t,t') = n_i V_i^2 \sum_{\vec p} G_0(\vec p;t,t'),
\end{align}
where $n_i$ and $V_i$ denote the impurity concentration and potential, respectively. 
For electron-phonon scattering in the Migdal limit of the Holstein model\cite{t_holstein_59},
we consider a single optical phonon mode
with frequency $\Omega$ coupled with a constant $g$, for which the self-energy is\cite{g_mahan}
\begin{align}
\Sigma(t,t') &= i g^2 D_0(t,t') \sum_{\vec p} G_0(\vec p; t,t') \\
D_0(t,t') &= -i \bigg[ \big[n_B(\Omega) + 1 - \theta_c(t,t')\big] e^{i\Omega(t-t')}
+ \big[ n_B(\Omega) + \theta_c(t,t') \big] e^{-i\Omega(t-t')} \bigg],
\end{align}
where $n_B(\Omega) = 1/(\exp(\Omega/T)-1)$.
The full Green's function is then self-consistently calculated through Dyson's equation expressed on the Keldysh contour as
\begin{align}
G(\vec k;t,t') &= G_0(\vec k;t,t') 
+ \ointclockwise dt_1 dt_2 G_0(\vec k;t,t_1) \Sigma(t_1,t_2) G(\vec k;t_2,t')
\end{align}
After discretization, this equation can be cast in the form of a matrix equation 
\begin{align}
\mathbf {G} = \mathbf G_0+ \mathbf{G_0} \cdot \left(\Delta_{t_1}\mathbf{ \Sigma} \Delta_{t_2}\right)\cdot \mathbf{G}
\end{align}
where all times including those of the differentials $\Delta_{t_{1,2}}$ lie on a branch of the Keldysh contour. This matrix equation can be solved using standard linear algebra techniques.

The process by which high-order harmonics are generated in undoped semiconducting crystals begins with tunnel ionization of electrons from the valence band into
the conduction band by the pump laser. Next, this same pulse accelerates the electrons, creating an oscillating current.
Here, the details of the tunnel ionization process are neglected and instead we focus on the motion of electrons only after they have been promoted to the conduction band. 
In principle, there is a time dependence due to the tunneling process since the tunneling rate
depends on the field, which is time dependent.  However, since the tunneling rate depends exponentially
on the field, the majority of electrons injected into the conduction band will tunnel there when the field is
largest, and will be essentially instantaneous. \cite{s_ghimire_11}
The subsequent driving force, in the absence of scattering
will give rise to a similar
harmonic spectrum as if the electrons had been there to begin with,
albeit with somewhat broader harmonics and some higher frequency content due to the nearly instantaneous
injection of the electrons into the valence band.
Finally, as pointed out by Golde et al.,\cite{d_golde_11} the interaction between the excited electrons
and the valence band holes leads to a contribution to the harmonic spectrum, in particular at
very high harmonic orders.  Here, we shall focus on the intra-band contribution, which is several orders
of magnitude stronger.

The instantaneous current $\vec j(t)$ created by the applied field is related to the lesser Green's function by
\begin{align}
\vec j(t) = -i\sum_{\vec k}\ \vec v(\vec k-\vec A(t)) \lim_{t'\rightarrow t} G^<(\vec k;t,t')
\end{align}
where $\vec v(\vec k) = \partial\epsilon(\vec k)/\partial \vec k\ $ is the group velocity of a
wave packet.    
We assume that the accelerated
electrons radiate proportional to the Fourier transform of the instantaneous current, with the HHG spectrum
proportional to
\begin{align}
\bigg| \int dt\ e^{i\omega t} \frac{d}{dt} \vec j(t)\bigg|^2 = \bigg|\omega \int dt\ e^{i\omega t} \vec j(t)\bigg|^2
\end{align}
We take the Fourier transform of the instantaneous current after interpolating with a shape-preserving Akima spline\cite{h_akima_70}. The self-energy and current are calculated on a $100\times100$ grid of momentum points in the two-dimensional square Brillouin zone, 
and $24\times24\times24$ grid in the three-dimensional Brillouin zone appropriate for the Si band structure, 
sufficient to give converged results for the current.

Since we are considering a laser pulse,  a DC contribution to the electric field
is excluded, as required by Maxwell's equations; any DC component does not propagate. Note that this restriction applies to the specific case of a propagating pulse; a DC contribution would be allowed for statically applied fields.
A corollary to the absence of a DC component of the field is that the current is also zero at
long times. This fact, combined with the relatively simple form of the current for a non-interacting system allows us to limit the length of the Keldysh contour to just the length of the pulse.
Our discretization on the real-time branches of the Keldysh contour varies depending on the section of the
paper, but is always kept above or equal to $\delta_t = 0.015 t_h^{-1}$, allowing up to the 30$^\textrm{th}$
harmonic to be visible (according to the Nyquist theorem)\cite{numerical_recipes}.
We always use a discretization of $10^{-2}\ T^{-1}$ along the imaginary-time spur, and set the
simulation temperature to 0.1 eV.
All times are measured in units of $t_h^{-1}$, and applied fields in units of $t_h$.
This can be converted to real units using appropriate values for the material under study.

For sample values of the hopping $t_h=1$ eV and material lattice constant or characteristic length scale $a_0=5$ \AA, a field strength of $E = t_h$ corresponds
to approximately 3.2 MV/cm (32 mV/\AA);
the temporal unit in this work, $t_h^{-1}$ would correspond to approximately 4.14 fs.
Note that these values are not uniformly applicable; $t_h$ and $a_0$ need to be adjusted for
the material under study.

\section{Scattering by impurities and phonons}
Motivated by the transition metal oxides,
we consider a square lattice with a model band structure (in the absence of an electric field)
given by
\begin{align}
\epsilon(\vec k) = -2 t_h (\cos(k_x) + \cos(k_y)) + 4 t^\prime_h \cos(k_x) \cos(k_y)-\mu \nonumber
\end{align}
where $t^\prime_h=0.3t_h$ and $\mu= -0.255t_h$.  We treat the system as isolated
\textemdash unconnected to a bath \textemdash so the chemical potential remains at the 
initial equilibrium value. 
The filling mainly affects the amplitude of the current, and does not affect qualitatively the results .
The system is driven by a Gaussian-shaped pump pulse centered at $t=0$
with a duration of $4.24 t_h^{-1}$, and 
a single underlying fundamental frequency $\omega_a=(9/5)\pi$ which corresponds to roughly 9 full cycles under the Gaussian envelope.
Since it is a pulse with finite width, the Fourier transform will contain a range of frequencies
centered around $\omega_a$, as will be discussed below.

\subsection{Harmonic generation from non-interacting electrons}

\begin{figure}[ht]
	\includegraphics[width=\columnwidth]{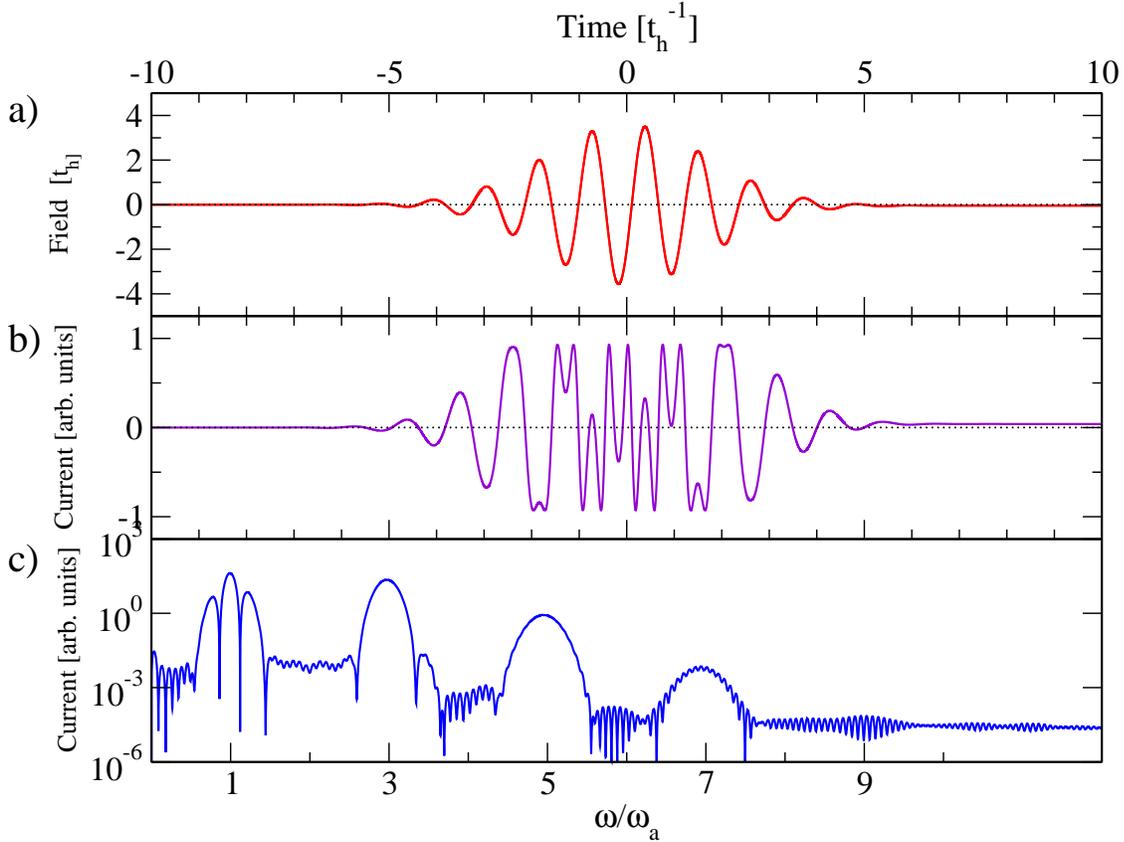}
	\caption{a) Field profile $A(t)$ for a multi-cycle Gaussian pulse with a maximum field amplitude of 
			\Emax$=20 t_h$.
		  (b) Instantaneous current response to this applied field. After the initial linear response, the field strength exceeds the point where Bloch oscillations occur, resulting in nonlinear behavior (see text).
		  (c) Semi-log plot of the HHG spectrum (see text), showing peaks at odd harmonic frequencies.}
	\label{fig:current}
\end{figure}
We calculate the current as a function of time for an oscillating field with a maximum field amplitude \Emax, as shown in Fig.~\ref{fig:current}. At the beginning of the pulse, the applied field amplitude is small and the field at that amplitude does not act very long; thus, the generated current is proportional to the field (up until $t \sim -2t_h^{-1}$), with equivalent current response exhibited at the end of the pulse. At larger amplitudes near the center of the pulse, the electrons are driven by the field for a sufficient period of time to hit the Brillouin zone boundary and Bragg scatter to the next zone,
where their velocity points in the opposite direction.

This phenomenon, commonly referred to as Bloch oscillations in condensed matter physics, originates 
from the periodicity of Bloch states. In the simple case of a constant DC field, the Peierls' substitution
\cite{r_peierls_33}
becomes
$\vec k\rightarrow\vec k + \vec E t$ and the $2\pi$ periodicity of the band structure 
means that the frequency of Bloch oscillations satisfies $\omega_B = 2\pi/E$. For oscillating fields, one can define an instantaneous Bloch frequency corresponding to the instantaneous field. Thus far, Bloch oscillations have not been observed directly in metals due to the 
experimental difficulty in resolving such short timescales and current degradation through impurity scattering, although oscillations have been observed in ultrasmall Josephson junctions,\cite{l_kuzmin_91} semiconductor superlattices,\cite{c_waschke_93,v_lyssenko_97} and optically trapped atoms.\cite{m_bendahan_96} However, the HHG spectrum is a direct consequence of the Bloch oscillations and provides an indirect measurement of them in solid-state systems, as discussed by various authors.\cite{d_golde_11,o_mucke_11,s_ghimire_11}
                                                                                                                               
As the field strength grows, the effects from Bloch oscillations intensify, as seen near the center of the pump pulse which displays several oscillations of the current in a region where the field does not change direction. 
A time-varying current due to the applied field leads to radiation, we thus
perform a Fourier transform of the current to obtain the HHG spectrum (shown in Fig.~\ref{fig:current}(c)).
\cite{s_ghimire_12}
On a semi-logarithmic scale, we clearly observe peaks near odd multiples of the central frequency of the applied field ($\omega_a$). The fact that odd multiples dominate
can be obtained from simple symmetry arguments and agrees with experimental observations. For inversion-symmetric systems,
the velocity is anti-symmetric under parity, which leads to a cancellation of even harmonics
of the current.\cite{f_faisal_97}
The peaks have a finite width resulting from the finite pulse length where as a continuous pulse would lead to sharp peaks at exactly odd harmonic frequencies.
\begin{figure}[ht]
	\includegraphics[clip=true,trim=20 40 80 40,width=0.9\columnwidth]{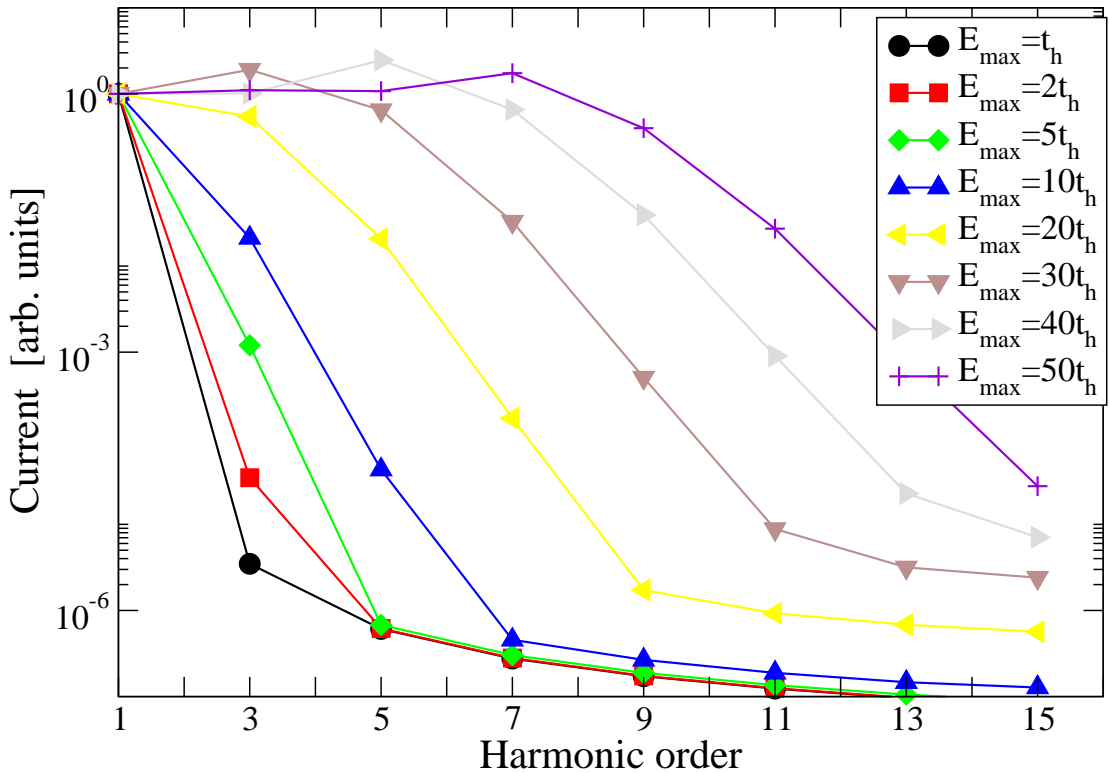}
	\caption{The HHG spectrum at odd harmonics of the driving frequency for various field amplitudes using the 2D square lattice tight-binding model.}
	\label{fig:longpulse_varyE}
\end{figure}

Figure~\ref{fig:longpulse_varyE} shows the HHG spectrum at integer multiples of the driving frequency for various field amplitudes. For the smallest fields, the majority of the electrons do not Bragg scatter from the Brillouin zone boundary and there are neither signatures of Bloch oscillations nor high-order harmonics. This can be seen in Figure~\ref{fig:longpulse_varyE} by comparing the relative heights of the $1^\textrm{st}$ and $3^\textrm{rd}$ harmonics; for the smallest field shown (\Emax$=t_h$) the ratio of the $1^\textrm{st}$ to the $3^\textrm{rd}$ harmonic is roughly $10^{-6}$; at this point, the current is entirely within the perturbative limit.
This ratio increases as a function of field with the largest fields showing a ratio approaching one.  
For the largest fields, the 1$^\mathrm{st}$ is not necessarily the largest, which can be understood from the time structure of the individual harmonics discussed below.

\subsection{Elastic scattering}

\begin{figure}[h]
	\includegraphics[trim=40 20 60 40, clip=true, width=0.8\columnwidth]{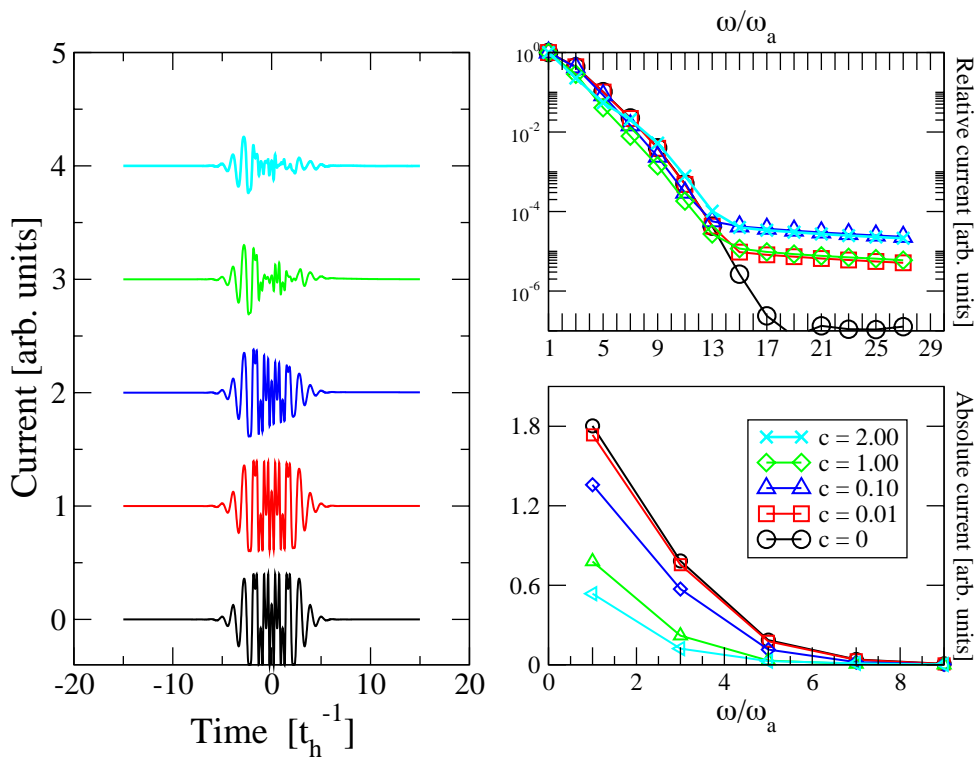}
	\caption{Left: Current response for various impurity strengths, $c = n_i V_i^2$, offset
		  for clarity. 
		  Right: Corresponding HHG spectra, absolute (bottom) and normalized to the 1$^\mathrm{st}$ harmonic (bottom)}
	\label{fig:imp_effects}
\end{figure}

To study simple effects from elastic scatterers on HHG, 
we consider a pump pulse whose field amplitude (\Amax=$20t_h$) produces a sufficient number of odd harmonics so that variations in the HHG spectrum as a function of impurity scattering strength can be tracked reasonably well. Figure~\ref{fig:imp_effects} shows the HHG spectra for increasing impurity strength,
where we define a measure of impurity strength as $c = n_i V_i^2$. As $c$ increases, the major effect is an overall decrease in the magnitude of the instantaneous current,
as can be seen in Fig.~\ref{fig:imp_effects}.
 In addition, near the center of the pulse where the Bloch oscillations are most noticeable, one observes a change in the current profile for the largest values of $c$.
The harmonic spectra are affected in a few ways.  Most noticeably, the ``noise floor'' where
individual harmonics can no longer be resolved (and the spectra go flat) lifts as the impurity scattering
increases.  This is due to the impurity scattering obscuring the smallest components of the electron
motion (corresponding to the highest harmonics), and thus raising the background noise.  Additionally,
any small features in the clean spectra due to harmonics of the band structure are
rapidly obscured.\cite{s_ghimire_11} 

\subsection{Inelastic scattering}
\begin{figure}[h]
	\includegraphics[trim=40 20 60 40, clip=true, width=0.8\columnwidth]{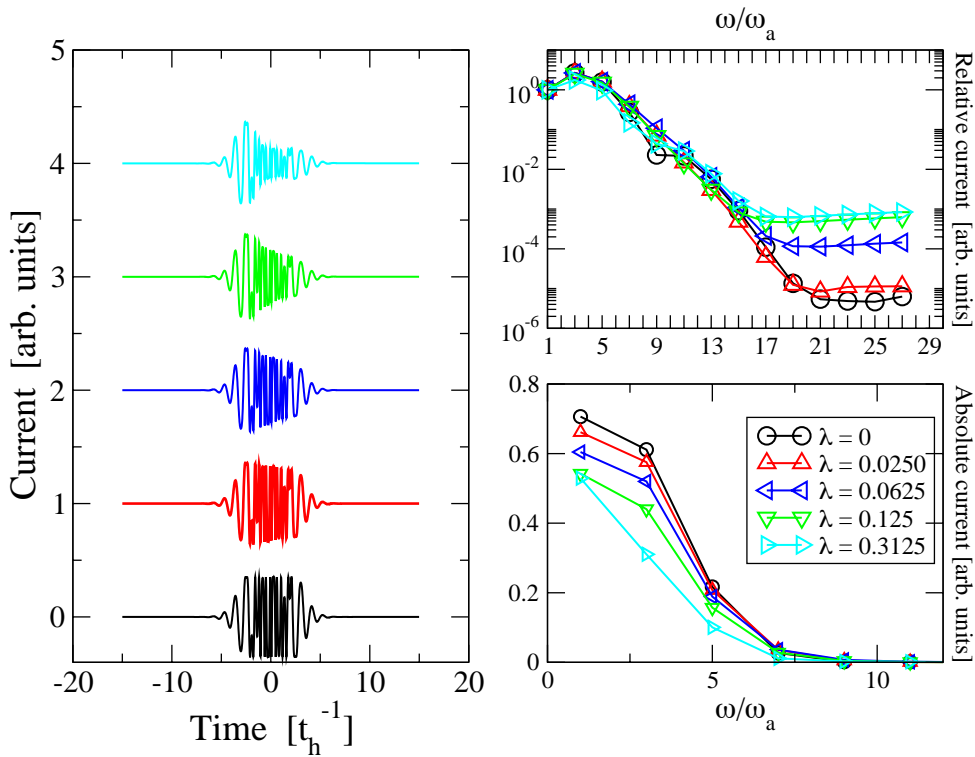}
	\caption{Left: Current response for various electron-phonon
	coupling strengths, offset
		  for clarity. 
		  Right: Corresponding HHG spectra, absolute (bottom) and normalized to the 1$^\mathrm{st}$ harmonic (bottom)}
	\label{fig:phonon_effects}
\end{figure}

In addition to impurity scattering, inelastic scattering can play a role in determining the current, and thus
the HHG spectra.
In particular, in semiconductors scattering due lattice vibrations (i.e. phonons) can play an
important role.\cite{v_zhukov_10}  Here, we consider the effect of a single phonon mode
on the generated spectra (as in the Holstein model).\cite{t_holstein_59}
Due to computational restrictions, the pulse strength is limited to \Amax=$5t_h$.
We use a phonon frequency of $\Omega = 0.2t_h$, and define the electron-phonon coupling
strength as $\lambda = 2 g^2 / \Omega W$, where $W$ is the bandwidth and $g$ is the
electron-phonon coupling constant.
With these parameters, we find the current and HHG spectra shown in Fig.~\ref{fig:phonon_effects}.
Similarly to the results for elastic scattering, although the overall current is strongly renormalized,
the scaled HHG spectra only differ qualitatively at the highest frequencies, where the individual
harmonics can no longer be resolved.

\section{Effects of Pulse Characteristics}
\begin{figure*}[ht]
	\includegraphics[clip=true,trim=35 25 35 90,width=0.49\textwidth]{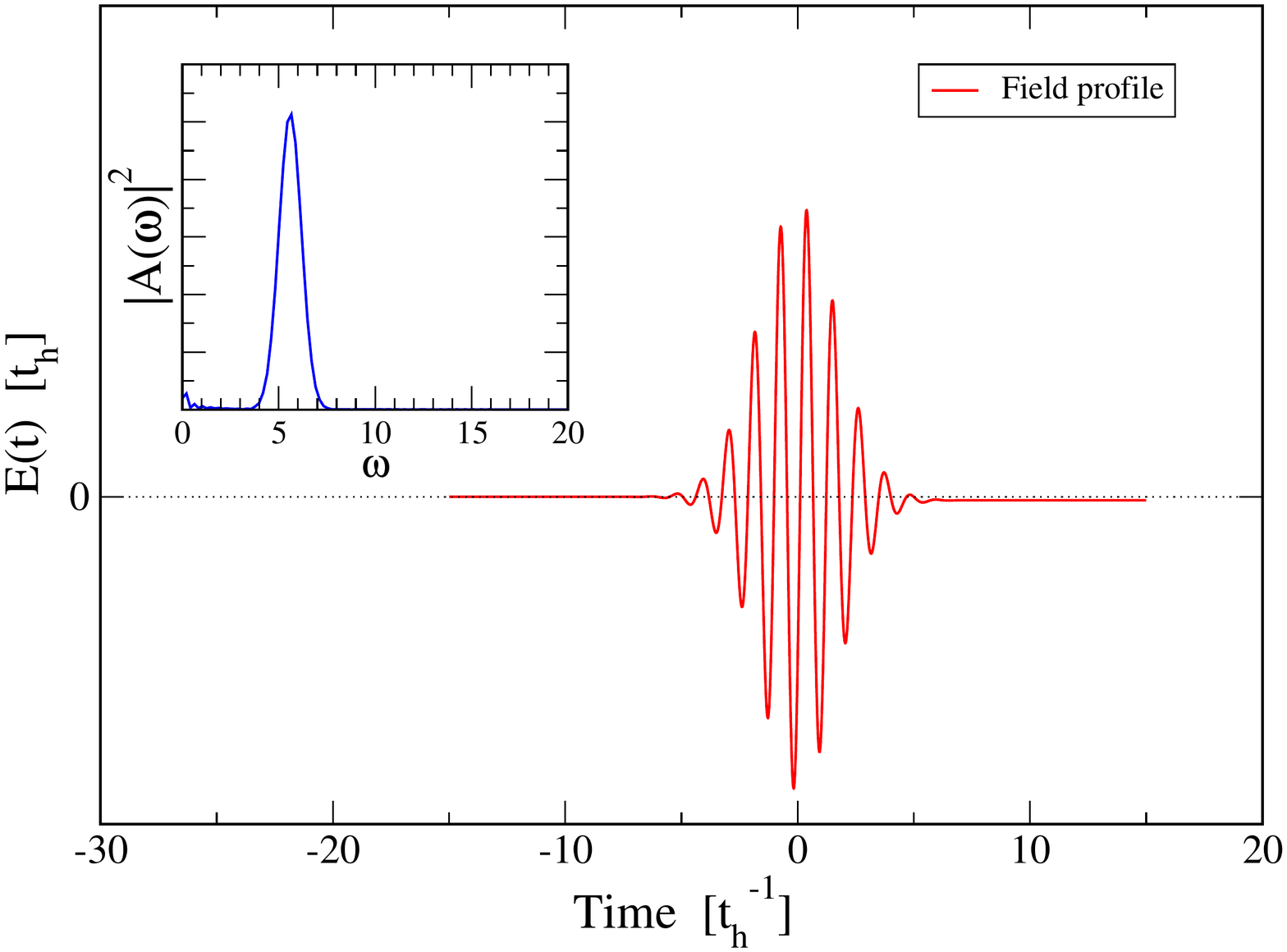}
	\includegraphics[clip=true,trim=35 25 35 90,width=0.49\textwidth]{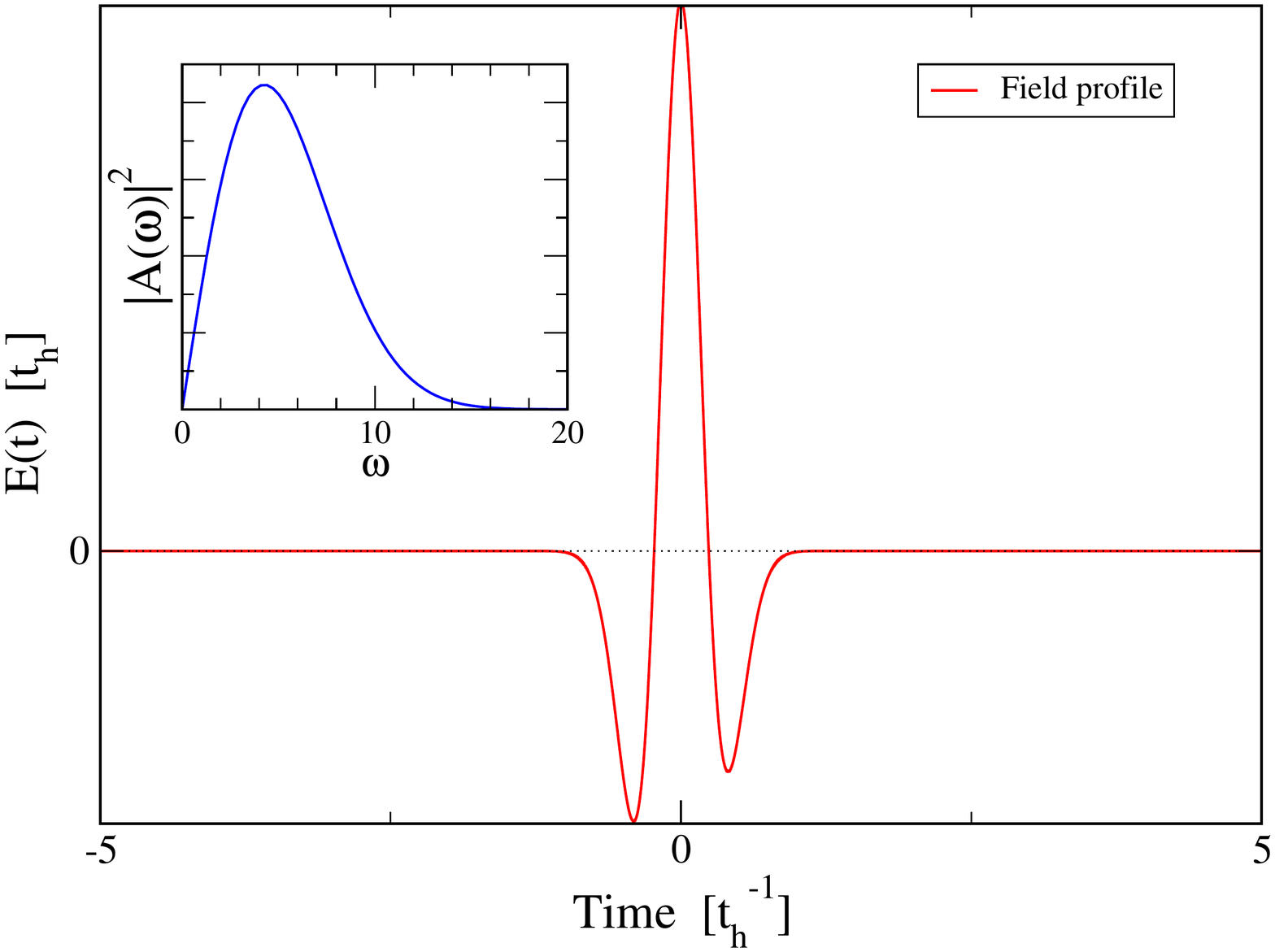}
	\caption{The field profile for the multi-cycle pulse (left) and single-cycle pulse (right). Inset: Fourier transform of the vector
		  potential (see text).}
	\label{fig:fieldprof}
\end{figure*}
\begin{figure*}[ht]
	\includegraphics[height=0.45\textwidth,angle=270]{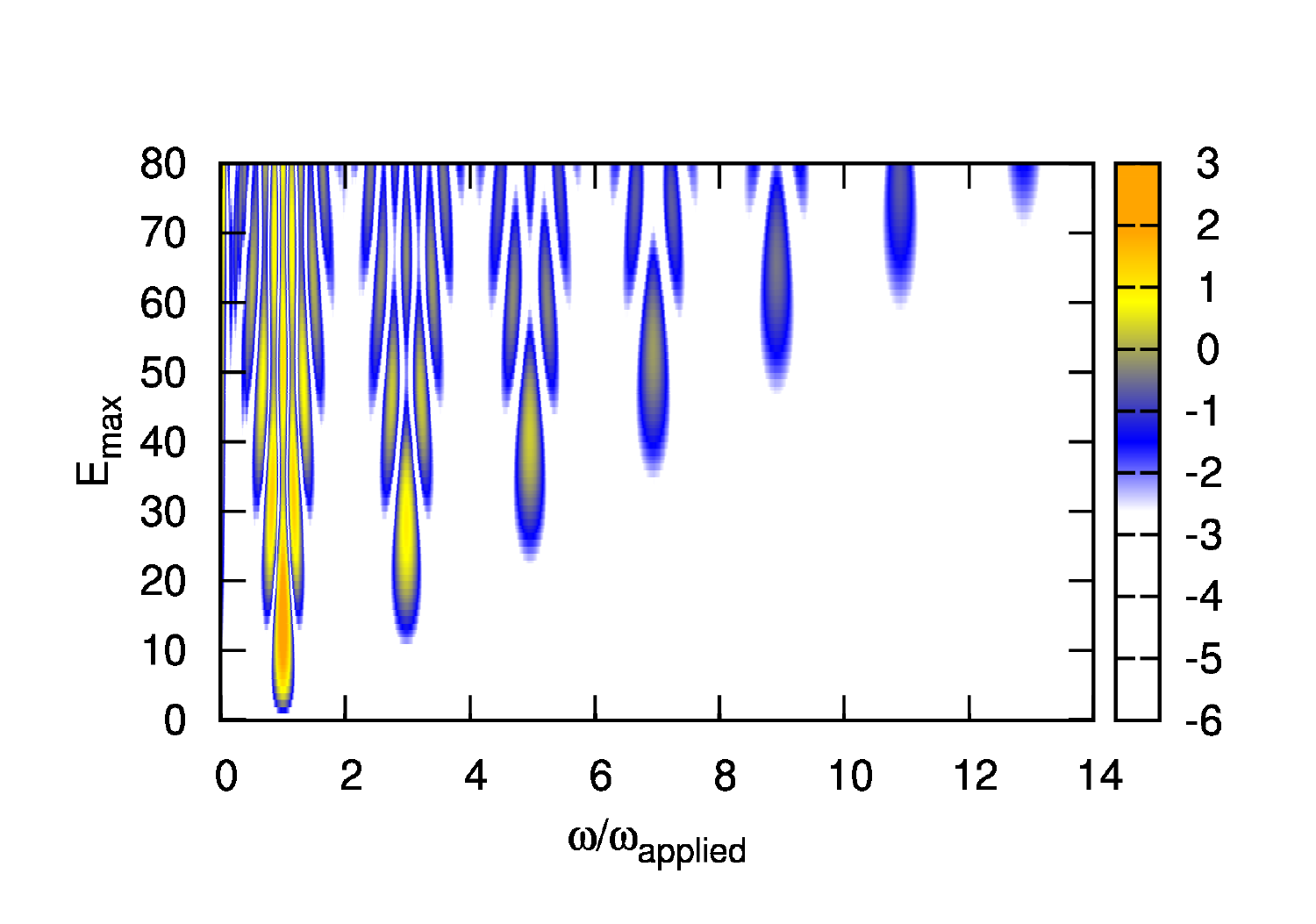}
	\includegraphics[height=0.45\textwidth,angle=270]{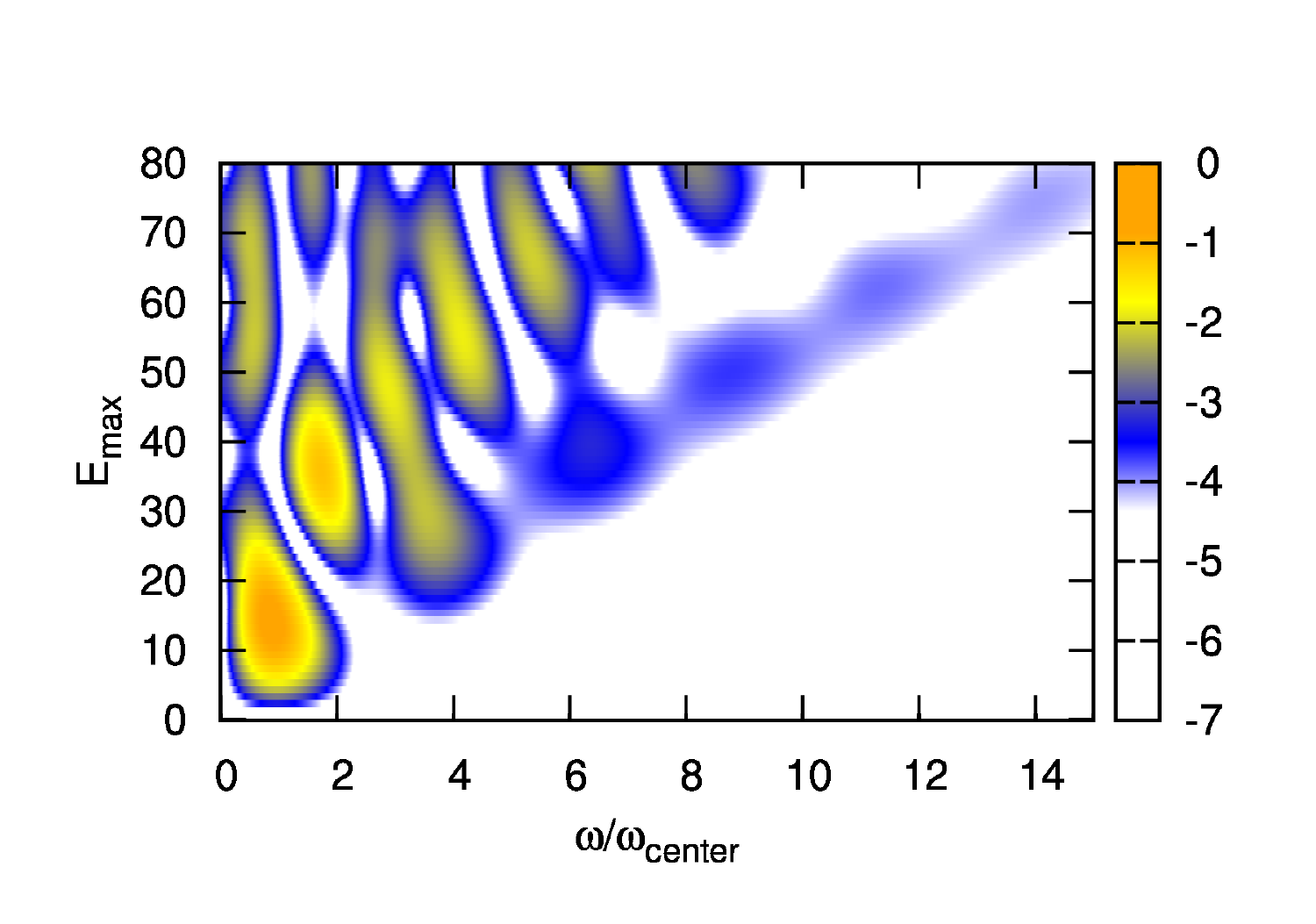}
	\caption{The HHG spectra for a multi-cycle (left) and single-cycle (right) pulse. The color
	scale is logarithmic in intensity. Please note that the horizontal axis is scaled by the
	central frequency of the pulse in both cases.}
	\label{fig:pulsecomp_ft_current}
\end{figure*}

The HHG spectra can be affected by the particulars of the applied pulse.  To illustrate this,
we consider two types of pulses: the multi-cycle pulse discussed above, 
and a single-cycle pulse similar to those used in THz experiments.\cite{a_lindenberg_priv_11}
For both pulses, although there is no DC component to the electric field,
there is still a choice of the carrier-envelope
phase.  The results here have an arbitrarily chosen phase; we present a detailed dependence of
the results of said phase in Section V.

The field profiles for the two pulses considered are shown in Fig.~\ref{fig:fieldprof}. The inset shows the Fourier transform of the vector potential obtained from the field profile. Due to the relatively large number of oscillations in the multi-cycle pulse, the Fourier
transform is strongly peaked at the driving frequency $\omega_a \approx 9\pi/5$. On the other hand, the single-cycle pulse is temporally short, leading to a large number of Fourier components with the largest contribution at $\omega_0\approx4t_h$. 
                                                                                                                                                                                                                                                                                                                                                                                                                                                                                                                 
Figure~\ref{fig:pulsecomp_ft_current} shows the HHG spectrum as a function of field amplitude and frequency. There is a clear distinction between the spectra associated with the multi-cycle and single-cycle pulses. As discussed above, the current induced by the multi-cycle pulse responds in odd harmonics of the fundamental frequency which is relatively well defined as seen in the inset of Fig.~\ref{fig:fieldprof}. However, for the single-cycle pulse, the response is more complex and we find two distinct regimes. When the field amplitude is small, the electronic system responds perturbatively as in the small amplitude regime for the multi-cycle pulse. As such, the dominant contribution to the HHG spectrum comes from that frequency which gives the largest contribution to the field, $\omega_0$. However, this linear response is limited to 
the fundamental, at a field amplitude where no harmonics can yet be observed. As the field grows larger, the electrons undergo Bragg reflection as discussed above, which leads to non-linear response in the current . Since the Fourier spectrum of the single-cycle pulse is very broad, the peaks in the Fourier-transformed current do not simply occur at odd 
harmonics of the fundamental applied frequency, but rather at frequencies that depend on the 
field strength and that are not linearly spaced. Additionally, the peaks observed are much broader than those of the multi-cycle pulse, which reflects the broad Fourier profile of the single-cycle pulse. Lastly, we note that the single-cycle pulse has a clear asymmetry about the pulse center. We have repeated the calculations with a symmetrized pulse, and find no qualitative difference in the HHG spectra.

\begin{figure}[ht]
	\includegraphics[width=0.49\textwidth,clip=true,trim=0 20 0 0]{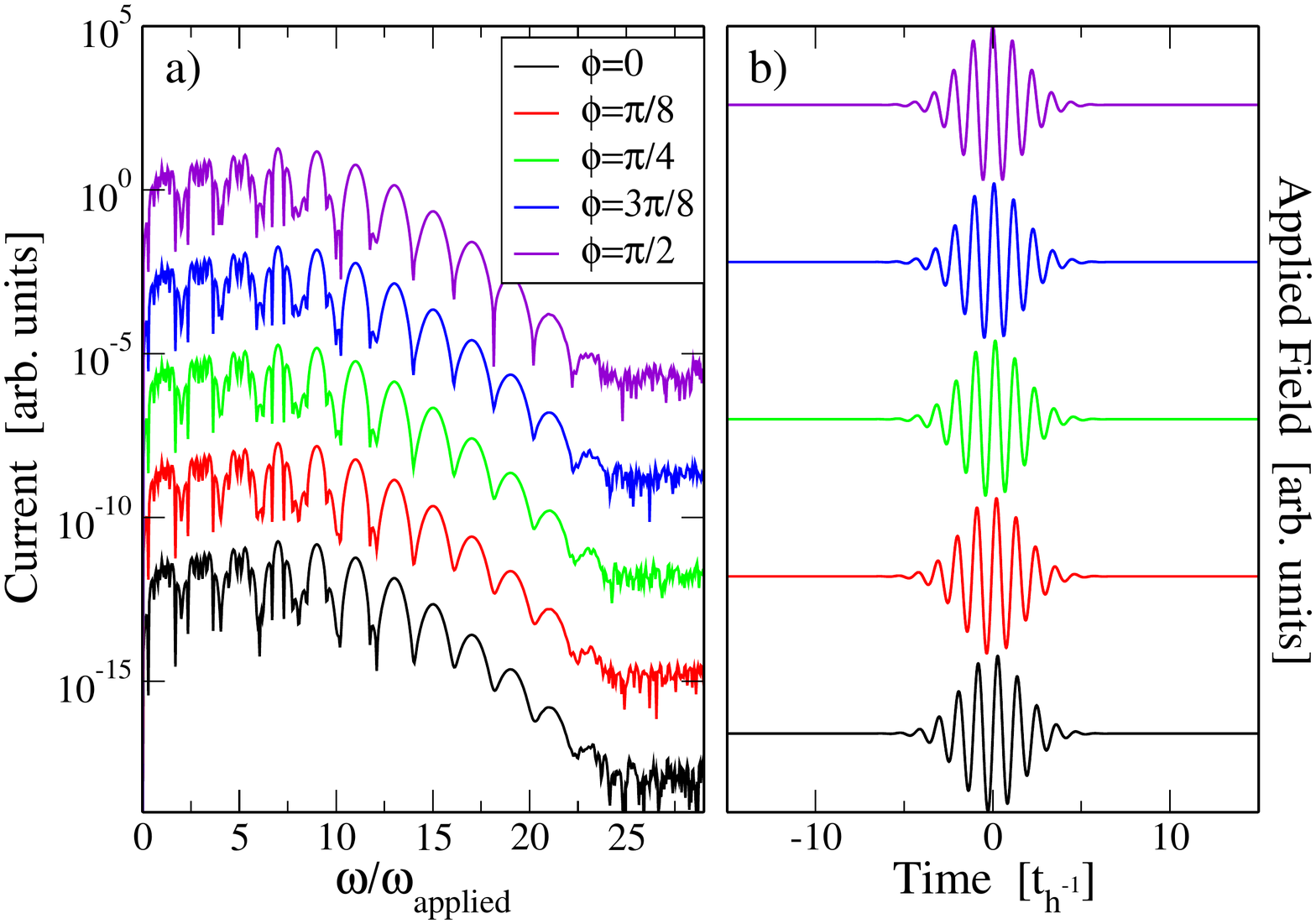}
	\includegraphics[width=0.49\textwidth,clip=true,trim=0 20 0 80]{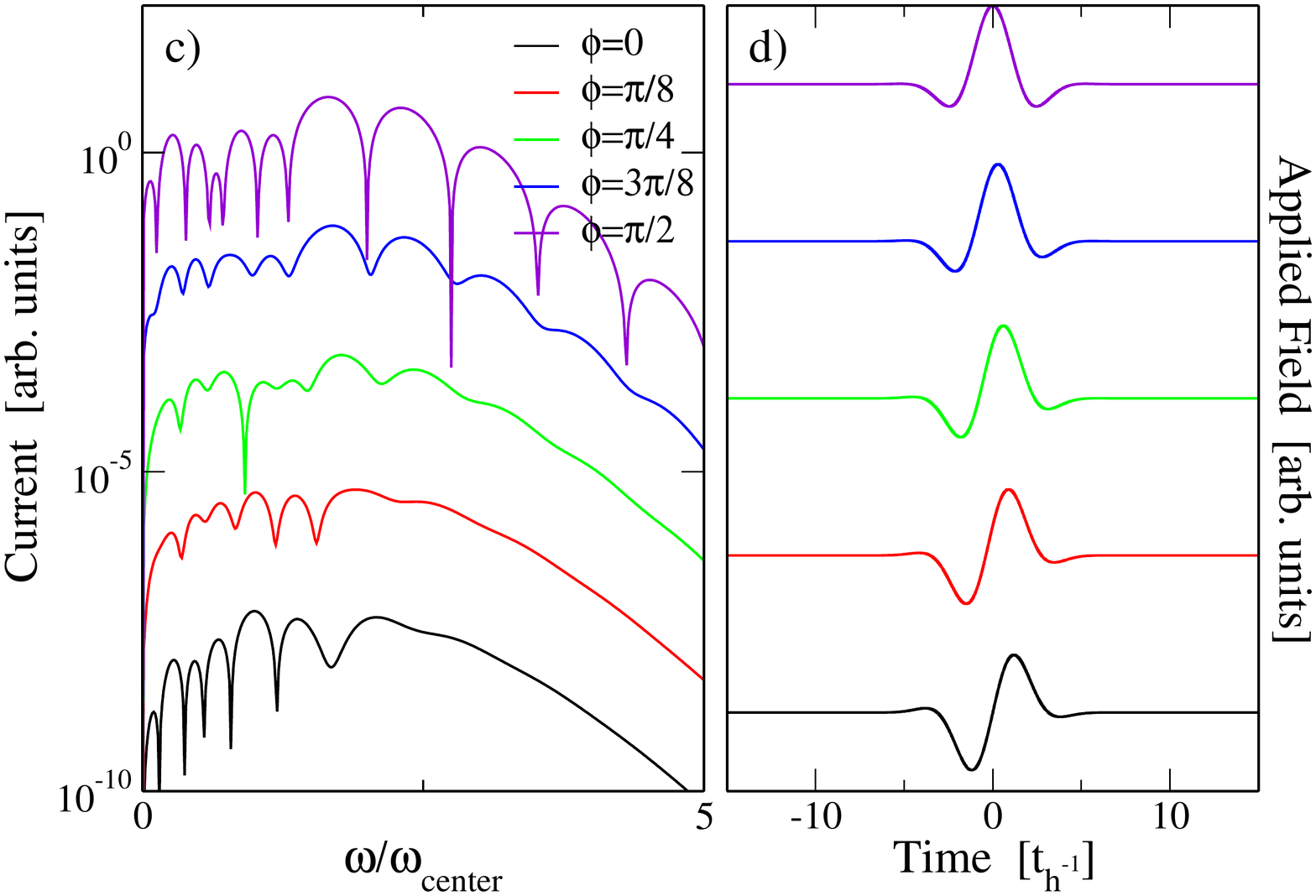}
	\caption{HHG as a function of carrier-envelope phase for (a) the 9-cycle and (c) single-cycle pulse.
	The HHG spectra are multiplied by a finite offset for clarity. Panels (b) and (d) show
	the vector potential $\vec A(t)$ for the 9-cycle and single-cycle pulses, respectively.}
	\label{fig:ce_phase}
\end{figure}
In addition to the field frequency and amplitude, it is possible to experimentally control
the carrier-envelope phase.  This is directly addressable within the formalism presented,
with the small caveat that there cannot be a DC component to the electric field.  Thus, we choose to
vary the phase $\phi$ in the vector potential:
\begin{align}
\vec A(t) = \vec A_{max} \sin(\omega t + \phi) \exp\left( -\frac{(t-t_0)^2}{2\sigma^2} \right).
\end{align}
Figure~\ref{fig:ce_phase} shows the variation in the HHG spectra due to the carrier-envelope phase
for both the 9-cycle and single-cycle pulse.  The development of
clear harmonic orders hinges on the applied field being sufficiently wide to clearly define
the carrier frequency.  Thus, the 9-cycle pulse does not show significant variation with the phase.
However, the Fourier transform of the single-cycle pulse is very wide; the pulse is not long enough for a
clear development of the carrier frequency. This causes a strong dependence on the carrier-envelope
phase.

\section{Time structure of the individual harmonics}
\begin{figure}[htpb]
	\includegraphics[width=\textwidth]{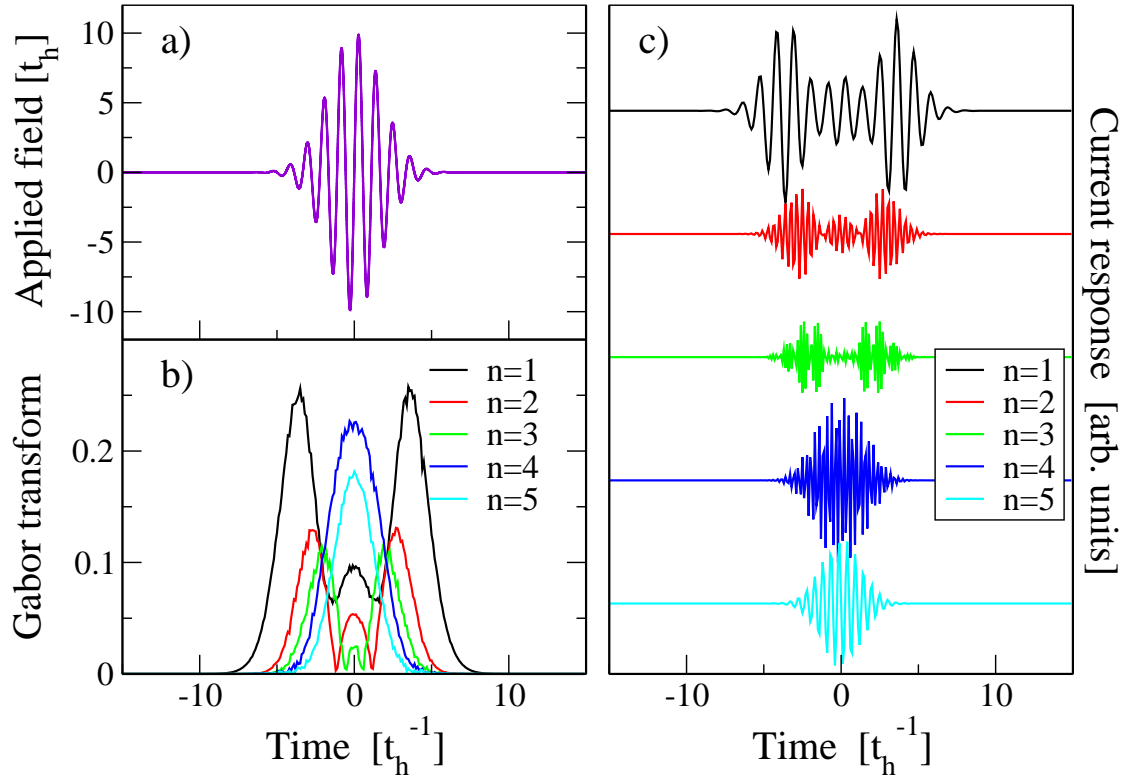}
	\caption{a)  Vector potential applied. b)  Windowed Fourier transform (Gabor transform)
	of the current, showing the power spectrum of the n$^\mathrm{th}$ harmonic as a function of time. 
	c) The individual harmonic contribution to the current.}
	\label{fig:gabor}
\end{figure}
The HHG process in solids
is due to the repeated Bragg scattering as the accelerated electrons hit the zone edge.This
leads to a simple time structure in the generated harmonics 
(see also Refs.~\onlinecite{o_mucke_11,s_ghimire_11,s_ghimire_12}). 
Figure~\ref{fig:gabor} shows the
windowed Fourier transform (also known as a Gabor transform\cite{gabor}) for the 9-cycle pulse. The low
order harmonics are mainly generated in the low-field regions of the pulse.  As the pulse grows
stronger, the low order harmonics are obscured by the higher orders, as the Bragg scattering occurs
more rapidly.  The higher orders are thus temporally localized to the center of the pulse.  This can be
clearly seen from both the time-dependent power spectra as well as the individual harmonics' contribution to the current.

Through temporal localization one can understand the increase of harmonic amplitudes above the
1$^\mathrm{st}$.
The low order harmonics are generated near the beginning and end of the pulse, and their duration is inversely related to the maximum field strength. Thus, for large field strengths, the time during which the 1$^\mathrm{st}$ order harmonic has a strong contribution is smaller than the next orders.

\section{Silicon Band Structure}

\begin{figure}[htpb]
	\includegraphics[trim=0 0 0 0, clip=true,width=0.75\textwidth]{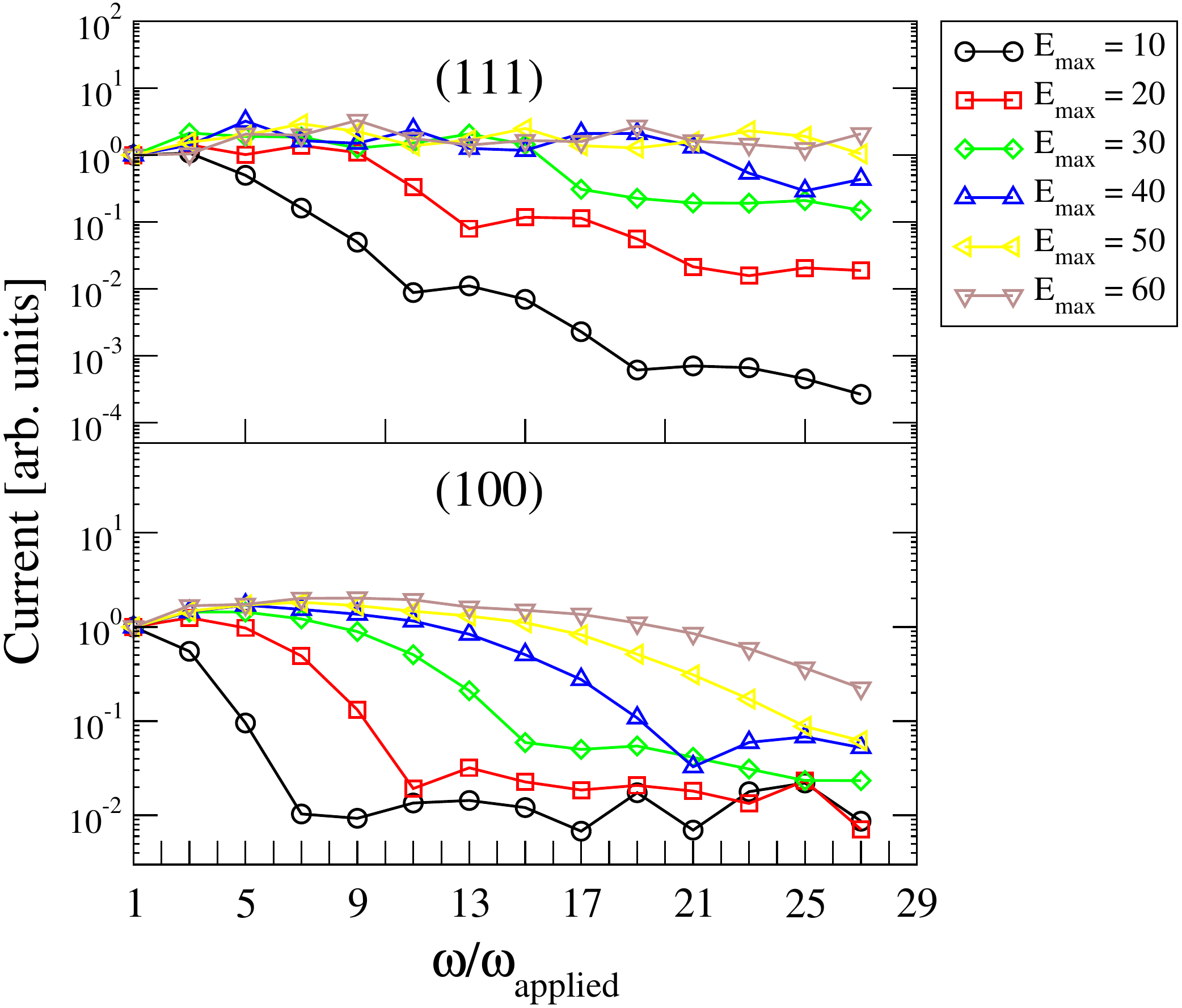}
	\caption{HHG spectra at integer multiples of the carrier frequency in Si for fields along
	(100) (top) and (111) (bottom), for varying field strengths. The units for the field are all in eV.}
	\label{fig:si_hhg}
\end{figure}
To investigate the effect of a realistic band structure on HHG, we have utilized a model for the conduction bands of Si as provided in some detail in Ref.~\onlinecite{p_vogl_83},
where we have placed the chemical potential just within the conduction bands (at the minimum
near the $X$ point).
Fig.~\ref{fig:si_hhg} shows the amplitude of the Fourier transformed current at odd multiples of $\omega_a$ (similar to the result shown in Fig.~\ref{fig:longpulse_varyE}). As in the case of the simple 2D band structure, the observed intensity increases with the field amplitude. 
Furthermore, since the bandwidth of the tight-binding model is different from the 2D model band
structure, the field strength at which the nonlinearity onsets is similarly different.
According to Ghimire et al.\cite{s_ghimire_11}, inclusion of more complex terms beyond first order in the band structure leads to an enhancement or reduction of the harmonic content in the HHG spectra.
Here we consider the multi-cycle pulse discussed above, applied along the (111) and (100) directions
of bulk Si. 
The band structure
in both directions is sufficiently complex that, after averaging over the
full Brillouin zone, there is no singular qualitative difference which can
be attributed to a particular feature of the band structure.  There is an overall difference
in the structure of the low  field harmonic spectra, 
which appear to decay faster in the (100) direction compared to the (111) direction.
The lack of a strong distinction in the spectra agrees with more detailed calculations done
based on density functional theory\cite{j_freericks_12}.

\section{Summary}
In summary, we have considered several aspects of harmonic generation in solids.  Harmonic generation
from solids shows promise as a future tool as a tunable light source for further experiments.  We find
that the harmonic spectra are sensitive to the particular details of the driving pulse, which indicates a
need to carefully control the parameters of said pulse.
On the  other hand,
the harmonic spectra are fairly robust against the influence of scattering, with the main
feature being a decrease in the overall amplitude of the current.  Finally, we find that for a
realistic band structure, the HHG spectra do not depend sensitively on direction due to the
complexity of the bands in any direction; however, this may not be the case in materials
with simpler band structures.
Overall, the most important message from this work is the very robust nature of
the HHG response in solids, which does not seem to depend very strongly
on the band structure, the orientation of the field, the scattering,
or the pulse characteristics, except in the special cases detailed above.

\acknowledgments{
AFK, BM and TPD were supported by the U.S. Department of Energy, 
Basic Energy Sciences, Materials Sciences and Engineering Division under 
contract No. DE-AC02-76SF00515. 
JKF was supported by the U.S. Department of Energy, 
Basic Energy Sciences, Materials Sciences and Engineering Division under contract No. 
DE-FG02-08ER46542 and by the McDevitt bequest at Georgetown University. 
The collaboration was supported by the U.S. Department of Energy, Basic Energy Sciences, 
Materials Sciences and Engineering Division under contract Nos. DE-FG02-08ER46540
and DE-SC0007091.
This work was made possible by the resources of the National Energy Research Scientific
Computing Center (via an Innovative and Novel Computational Impact on Theory and Experiment
grant) which is supported by the U.S. DOE, Office of Science, under Contract No. DE-AC02-05CH11231.
We gratefully acknowledge discussions with P.~S. Kirchmann, A.~M. Lindenberg and 
D.~A. Reis.}

\bibliography{timedomain}

\begin{thebibliography}{53}
\expandafter\ifx\csname natexlab\endcsname\relax\def\natexlab#1{#1}\fi
\expandafter\ifx\csname bibnamefont\endcsname\relax
  \def\bibnamefont#1{#1}\fi
\expandafter\ifx\csname bibfnamefont\endcsname\relax
  \def\bibfnamefont#1{#1}\fi
\expandafter\ifx\csname citenamefont\endcsname\relax
  \def\citenamefont#1{#1}\fi
\expandafter\ifx\csname url\endcsname\relax
  \def\url#1{\texttt{#1}}\fi
\expandafter\ifx\csname urlprefix\endcsname\relax\def\urlprefix{URL }\fi
\providecommand{\bibinfo}[2]{#2}
\providecommand{\eprint}[2][]{\url{#2}}

\bibitem[{\citenamefont{{Cavalieri} et~al.}(2007)\citenamefont{{Cavalieri},
  {M{\"u}ller}, {Uphues}, {Yakovlev}, {Baltu{\v s}ka}, {Horvath}, {Schmidt},
  {Bl{\"u}mel}, {Holzwarth}, {Hendel} et~al.}}]{a_cavalieri_07}
\bibinfo{author}{\bibfnamefont{A.~L.} \bibnamefont{{Cavalieri}}},
  \bibinfo{author}{\bibfnamefont{N.}~\bibnamefont{{M{\"u}ller}}},
  \bibinfo{author}{\bibfnamefont{T.}~\bibnamefont{{Uphues}}},
  \bibinfo{author}{\bibfnamefont{V.~S.} \bibnamefont{{Yakovlev}}},
  \bibinfo{author}{\bibfnamefont{A.}~\bibnamefont{{Baltu{\v s}ka}}},
  \bibinfo{author}{\bibfnamefont{B.}~\bibnamefont{{Horvath}}},
  \bibinfo{author}{\bibfnamefont{B.}~\bibnamefont{{Schmidt}}},
  \bibinfo{author}{\bibfnamefont{L.}~\bibnamefont{{Bl{\"u}mel}}},
  \bibinfo{author}{\bibfnamefont{R.}~\bibnamefont{{Holzwarth}}},
  \bibinfo{author}{\bibfnamefont{S.}~\bibnamefont{{Hendel}}},
  \bibnamefont{et~al.}, \bibinfo{journal}{Nature}
  \textbf{\bibinfo{volume}{449}}, \bibinfo{pages}{1029} (\bibinfo{year}{2007}).

\bibitem[{\citenamefont{Schmitt et~al.}(2008)\citenamefont{Schmitt, Kirchmann,
  Bovensiepen, Moore, Rettig, Krenz, Chu, Ru, Perfetti, Lu
  et~al.}}]{f_schmitt_08}
\bibinfo{author}{\bibfnamefont{F.}~\bibnamefont{Schmitt}},
  \bibinfo{author}{\bibfnamefont{P.~S.} \bibnamefont{Kirchmann}},
  \bibinfo{author}{\bibfnamefont{U.}~\bibnamefont{Bovensiepen}},
  \bibinfo{author}{\bibfnamefont{R.~G.} \bibnamefont{Moore}},
  \bibinfo{author}{\bibfnamefont{L.}~\bibnamefont{Rettig}},
  \bibinfo{author}{\bibfnamefont{M.}~\bibnamefont{Krenz}},
  \bibinfo{author}{\bibfnamefont{J.~H.} \bibnamefont{Chu}},
  \bibinfo{author}{\bibfnamefont{N.}~\bibnamefont{Ru}},
  \bibinfo{author}{\bibfnamefont{L.}~\bibnamefont{Perfetti}},
  \bibinfo{author}{\bibfnamefont{D.~H.} \bibnamefont{Lu}},
  \bibnamefont{et~al.}, \bibinfo{journal}{Science}
  \textbf{\bibinfo{volume}{321}}, \bibinfo{pages}{1649} (\bibinfo{year}{2008}).

\bibitem[{\citenamefont{Perfetti et~al.}(2008)\citenamefont{Perfetti, Loukakos,
  Lisowski, Bovensiepen, Wolf, Berger, Biermann, and Georges}}]{l_perfetti_08}
\bibinfo{author}{\bibfnamefont{L.}~\bibnamefont{Perfetti}},
  \bibinfo{author}{\bibfnamefont{P.~A.} \bibnamefont{Loukakos}},
  \bibinfo{author}{\bibfnamefont{M.}~\bibnamefont{Lisowski}},
  \bibinfo{author}{\bibfnamefont{U.}~\bibnamefont{Bovensiepen}},
  \bibinfo{author}{\bibfnamefont{M.}~\bibnamefont{Wolf}},
  \bibinfo{author}{\bibfnamefont{H.}~\bibnamefont{Berger}},
  \bibinfo{author}{\bibfnamefont{S.}~\bibnamefont{Biermann}}, \bibnamefont{and}
  \bibinfo{author}{\bibfnamefont{A.}~\bibnamefont{Georges}},
  \bibinfo{journal}{New. J. Phys.} \textbf{\bibinfo{volume}{10}},
  \bibinfo{pages}{053019} (\bibinfo{year}{2008}).

\bibitem[{\citenamefont{Basov et~al.}(2011)\citenamefont{Basov, Averitt,
  van~der Marel, Dressel, and Haule}}]{d_basov_11}
\bibinfo{author}{\bibfnamefont{D.~N.} \bibnamefont{Basov}},
  \bibinfo{author}{\bibfnamefont{R.~D.} \bibnamefont{Averitt}},
  \bibinfo{author}{\bibfnamefont{D.}~\bibnamefont{van~der Marel}},
  \bibinfo{author}{\bibfnamefont{M.}~\bibnamefont{Dressel}}, \bibnamefont{and}
  \bibinfo{author}{\bibfnamefont{K.}~\bibnamefont{Haule}},
  \bibinfo{journal}{Rev. Mod. Phys.} \textbf{\bibinfo{volume}{83}},
  \bibinfo{pages}{471} (\bibinfo{year}{2011}).

\bibitem[{\citenamefont{Cavalleri et~al.}(2001)\citenamefont{Cavalleri,
  T{\'o}th, Siders, Squier, R{\'a}ksi, Forget, and Kieffer}}]{a_cavalleri_01}
\bibinfo{author}{\bibfnamefont{A.}~\bibnamefont{Cavalleri}},
  \bibinfo{author}{\bibfnamefont{C.}~\bibnamefont{T{\'o}th}},
  \bibinfo{author}{\bibfnamefont{C.}~\bibnamefont{Siders}},
  \bibinfo{author}{\bibfnamefont{J.}~\bibnamefont{Squier}},
  \bibinfo{author}{\bibfnamefont{F.}~\bibnamefont{R{\'a}ksi}},
  \bibinfo{author}{\bibfnamefont{P.}~\bibnamefont{Forget}}, \bibnamefont{and}
  \bibinfo{author}{\bibfnamefont{J.}~\bibnamefont{Kieffer}},
  \bibinfo{journal}{\prl} \textbf{\bibinfo{volume}{87}} (\bibinfo{year}{2001}).

\bibitem[{\citenamefont{Rousse et~al.}(2001)\citenamefont{Rousse, Rischel, and
  Gauthier}}]{a_rousse_01}
\bibinfo{author}{\bibfnamefont{A.}~\bibnamefont{Rousse}},
  \bibinfo{author}{\bibfnamefont{C.}~\bibnamefont{Rischel}}, \bibnamefont{and}
  \bibinfo{author}{\bibfnamefont{J.-C.} \bibnamefont{Gauthier}},
  \bibinfo{journal}{Rev. Mod. Phys.} \textbf{\bibinfo{volume}{73}},
  \bibinfo{pages}{17} (\bibinfo{year}{2001}).

\bibitem[{\citenamefont{Tritschler et~al.}(2003)\citenamefont{Tritschler,
  M\"ucke, and Wegener}}]{t_tritschler_03}
\bibinfo{author}{\bibfnamefont{T.}~\bibnamefont{Tritschler}},
  \bibinfo{author}{\bibfnamefont{O.~D.} \bibnamefont{M\"ucke}},
  \bibnamefont{and} \bibinfo{author}{\bibfnamefont{M.}~\bibnamefont{Wegener}},
  \bibinfo{journal}{Phys. Rev. A} \textbf{\bibinfo{volume}{68}},
  \bibinfo{pages}{033404} (\bibinfo{year}{2003}).

\bibitem[{\citenamefont{Golde et~al.}(2008)\citenamefont{Golde, Meier, and
  Koch}}]{d_golde_08}
\bibinfo{author}{\bibfnamefont{D.}~\bibnamefont{Golde}},
  \bibinfo{author}{\bibfnamefont{T.}~\bibnamefont{Meier}}, \bibnamefont{and}
  \bibinfo{author}{\bibfnamefont{S.~W.} \bibnamefont{Koch}},
  \bibinfo{journal}{Phys. Rev. B} \textbf{\bibinfo{volume}{77}},
  \bibinfo{pages}{075330} (\bibinfo{year}{2008}).

\bibitem[{\citenamefont{Golde et~al.}(2011)\citenamefont{Golde, Kira, Meier,
  and Koch}}]{d_golde_11}
\bibinfo{author}{\bibfnamefont{D.}~\bibnamefont{Golde}},
  \bibinfo{author}{\bibfnamefont{M.}~\bibnamefont{Kira}},
  \bibinfo{author}{\bibfnamefont{T.}~\bibnamefont{Meier}}, \bibnamefont{and}
  \bibinfo{author}{\bibfnamefont{S.~W.} \bibnamefont{Koch}},
  \bibinfo{journal}{Physica Status Solidi (b)} \textbf{\bibinfo{volume}{248}},
  \bibinfo{pages}{863} (\bibinfo{year}{2011}).

\bibitem[{\citenamefont{M\"ucke}(2011)}]{o_mucke_11}
\bibinfo{author}{\bibfnamefont{O.~D.} \bibnamefont{M\"ucke}},
  \bibinfo{journal}{Phys. Rev. B} \textbf{\bibinfo{volume}{84}},
  \bibinfo{pages}{081202(R)} (\bibinfo{year}{2011}).

\bibitem[{\citenamefont{Faisal and Kami\'nski}(1997)}]{f_faisal_97}
\bibinfo{author}{\bibfnamefont{F.~H.~M.} \bibnamefont{Faisal}}
  \bibnamefont{and} \bibinfo{author}{\bibfnamefont{J.~Z.}
  \bibnamefont{Kami\'nski}}, \bibinfo{journal}{Phys. Rev. A}
  \textbf{\bibinfo{volume}{56}}, \bibinfo{pages}{748} (\bibinfo{year}{1997}).

\bibitem[{\citenamefont{DeVries}(1990)}]{p_devries_90}
\bibinfo{author}{\bibfnamefont{P.~L.} \bibnamefont{DeVries}},
  \bibinfo{journal}{J. Opt. Soc. Am. B} \textbf{\bibinfo{volume}{7}},
  \bibinfo{pages}{517} (\bibinfo{year}{1990}).

\bibitem[{\citenamefont{Kadanoff and Baym}(1962)}]{kadanoff_baym}
\bibinfo{author}{\bibfnamefont{L.~P.} \bibnamefont{Kadanoff}} \bibnamefont{and}
  \bibinfo{author}{\bibfnamefont{G.}~\bibnamefont{Baym}},
  \emph{\bibinfo{title}{Quantum Statistical Mechanics}} (\bibinfo{publisher}{W.
  A. Benjamin, Inc., NY}, \bibinfo{year}{1962}).

\bibitem[{\citenamefont{{Keldysh}}(1964)}]{l_keldysh_64}
\bibinfo{author}{\bibfnamefont{L.~V.} \bibnamefont{{Keldysh}}},
  \bibinfo{journal}{Zh. Eksp. Teor. Fiz.} \textbf{\bibinfo{volume}{47}},
  \bibinfo{pages}{1515} (\bibinfo{year}{1964}).

\bibitem[{\citenamefont{Turkowski and Freericks}(2005)}]{v_turkowski_05}
\bibinfo{author}{\bibfnamefont{V.}~\bibnamefont{Turkowski}} \bibnamefont{and}
  \bibinfo{author}{\bibfnamefont{J.~K.} \bibnamefont{Freericks}},
  \bibinfo{journal}{\prb} \textbf{\bibinfo{volume}{71}}, \bibinfo{pages}{85104}
  (\bibinfo{year}{2005}).

\bibitem[{\citenamefont{Freericks et~al.}(2006)\citenamefont{Freericks,
  Turkowski, and Zlatic'}}]{j_freericks_06}
\bibinfo{author}{\bibfnamefont{J.~K.} \bibnamefont{Freericks}},
  \bibinfo{author}{\bibfnamefont{V.}~\bibnamefont{Turkowski}},
  \bibnamefont{and} \bibinfo{author}{\bibfnamefont{V.}~\bibnamefont{Zlatic'}},
  \bibinfo{journal}{\prl} \textbf{\bibinfo{volume}{97}},
  \bibinfo{pages}{266408} (\bibinfo{year}{2006}).

\bibitem[{\citenamefont{{Eckstein} and {Kollar}}(2008)}]{m_eckstein_08}
\bibinfo{author}{\bibfnamefont{M.}~\bibnamefont{{Eckstein}}} \bibnamefont{and}
  \bibinfo{author}{\bibfnamefont{M.}~\bibnamefont{{Kollar}}},
  \bibinfo{journal}{\prl} \textbf{\bibinfo{volume}{100}},
  \bibinfo{pages}{120404} (\bibinfo{year}{2008}).

\bibitem[{\citenamefont{{Freericks}}(2008)}]{j_freericks_08}
\bibinfo{author}{\bibfnamefont{J.~K.} \bibnamefont{{Freericks}}},
  \bibinfo{journal}{\prb} \textbf{\bibinfo{volume}{77}},
  \bibinfo{pages}{075109} (\bibinfo{year}{2008}).

\bibitem[{\citenamefont{{Eckstein} et~al.}(2009)\citenamefont{{Eckstein},
  {Hackl}, {Kehrein}, {Kollar}, {Moeckel}, {Werner}, and
  {Wolf}}}]{m_eckstein_09}
\bibinfo{author}{\bibfnamefont{M.}~\bibnamefont{{Eckstein}}},
  \bibinfo{author}{\bibfnamefont{A.}~\bibnamefont{{Hackl}}},
  \bibinfo{author}{\bibfnamefont{S.}~\bibnamefont{{Kehrein}}},
  \bibinfo{author}{\bibfnamefont{M.}~\bibnamefont{{Kollar}}},
  \bibinfo{author}{\bibfnamefont{M.}~\bibnamefont{{Moeckel}}},
  \bibinfo{author}{\bibfnamefont{P.}~\bibnamefont{{Werner}}}, \bibnamefont{and}
  \bibinfo{author}{\bibfnamefont{F.~A.} \bibnamefont{{Wolf}}},
  \bibinfo{journal}{Eur. Phys. J.: S.T.} \textbf{\bibinfo{volume}{180}},
  \bibinfo{pages}{217} (\bibinfo{year}{2009}).

\bibitem[{\citenamefont{{Moritz} et~al.}(2010)\citenamefont{{Moritz},
  {Devereaux}, and {Freericks}}}]{b_moritz_10}
\bibinfo{author}{\bibfnamefont{B.}~\bibnamefont{{Moritz}}},
  \bibinfo{author}{\bibfnamefont{T.~P.} \bibnamefont{{Devereaux}}},
  \bibnamefont{and} \bibinfo{author}{\bibfnamefont{J.~K.}
  \bibnamefont{{Freericks}}}, \bibinfo{journal}{\prb}
  \textbf{\bibinfo{volume}{81}}, \bibinfo{pages}{165112}
  (\bibinfo{year}{2010}).

\bibitem[{\citenamefont{{Eckstein} et~al.}(2010)\citenamefont{{Eckstein},
  {Kollar}, and {Werner}}}]{m_eckstein_10}
\bibinfo{author}{\bibfnamefont{M.}~\bibnamefont{{Eckstein}}},
  \bibinfo{author}{\bibfnamefont{M.}~\bibnamefont{{Kollar}}}, \bibnamefont{and}
  \bibinfo{author}{\bibfnamefont{P.}~\bibnamefont{{Werner}}},
  \bibinfo{journal}{\prb} \textbf{\bibinfo{volume}{81}},
  \bibinfo{pages}{115131} (\bibinfo{year}{2010}).

\bibitem[{\citenamefont{{Moritz} et~al.}(2011)\citenamefont{{Moritz},
  {Devereaux}, and {Freericks}}}]{b_moritz_11}
\bibinfo{author}{\bibfnamefont{B.}~\bibnamefont{{Moritz}}},
  \bibinfo{author}{\bibfnamefont{T.~P.} \bibnamefont{{Devereaux}}},
  \bibnamefont{and} \bibinfo{author}{\bibfnamefont{J.~K.}
  \bibnamefont{{Freericks}}}, \bibinfo{journal}{Comp. Phys. Comm.}
  \textbf{\bibinfo{volume}{182}}, \bibinfo{pages}{109} (\bibinfo{year}{2011}).

\bibitem[{\citenamefont{Eckstein et~al.}(2009)\citenamefont{Eckstein, Kollar,
  and Werner}}]{m_eckstein_09b}
\bibinfo{author}{\bibfnamefont{M.}~\bibnamefont{Eckstein}},
  \bibinfo{author}{\bibfnamefont{M.}~\bibnamefont{Kollar}}, \bibnamefont{and}
  \bibinfo{author}{\bibfnamefont{P.}~\bibnamefont{Werner}},
  \bibinfo{journal}{\prl} \textbf{\bibinfo{volume}{103}},
  \bibinfo{pages}{056403} (\bibinfo{year}{2009}).

\bibitem[{\citenamefont{{Balzer} and {Potthoff}}(2011)}]{m_balzer_11}
\bibinfo{author}{\bibfnamefont{M.}~\bibnamefont{{Balzer}}} \bibnamefont{and}
  \bibinfo{author}{\bibfnamefont{M.}~\bibnamefont{{Potthoff}}},
  \bibinfo{journal}{\prb} \textbf{\bibinfo{volume}{83}},
  \bibinfo{pages}{195132} (\bibinfo{year}{2011}).

\bibitem[{\citenamefont{Park and Light}(1986)}]{t_park_86}
\bibinfo{author}{\bibfnamefont{T.~J.} \bibnamefont{Park}} \bibnamefont{and}
  \bibinfo{author}{\bibfnamefont{J.~C.} \bibnamefont{Light}},
  \bibinfo{journal}{J. Chem. Phys.} \textbf{\bibinfo{volume}{85}},
  \bibinfo{pages}{5870} (\bibinfo{year}{1986}).

\bibitem[{\citenamefont{Anders and Schiller}(2005)}]{f_anders_05}
\bibinfo{author}{\bibfnamefont{F.}~\bibnamefont{Anders}} \bibnamefont{and}
  \bibinfo{author}{\bibfnamefont{A.}~\bibnamefont{Schiller}},
  \bibinfo{journal}{\prl} \textbf{\bibinfo{volume}{95}},
  \bibinfo{pages}{196801} (\bibinfo{year}{2005}).

\bibitem[{\citenamefont{White and Feiguin}(2004)}]{s_white_04}
\bibinfo{author}{\bibfnamefont{S.~R.} \bibnamefont{White}} \bibnamefont{and}
  \bibinfo{author}{\bibfnamefont{A.~E.} \bibnamefont{Feiguin}},
  \bibinfo{journal}{\prl} \textbf{\bibinfo{volume}{93}},
  \bibinfo{pages}{076401} (\bibinfo{year}{2004}).

\bibitem[{\citenamefont{Alvarez et~al.}(2011)\citenamefont{Alvarez, Dias~da
  Silva, Ponce, and Dagotto}}]{g_alvarez_11}
\bibinfo{author}{\bibfnamefont{G.}~\bibnamefont{Alvarez}},
  \bibinfo{author}{\bibfnamefont{L.~G. G.~V.} \bibnamefont{Dias~da Silva}},
  \bibinfo{author}{\bibfnamefont{E.}~\bibnamefont{Ponce}}, \bibnamefont{and}
  \bibinfo{author}{\bibfnamefont{E.}~\bibnamefont{Dagotto}},
  \bibinfo{journal}{Phys. Rev. E} \textbf{\bibinfo{volume}{84}},
  \bibinfo{pages}{056706} (\bibinfo{year}{2011}).

\bibitem[{\citenamefont{Antoine et~al.}(1996)\citenamefont{Antoine, L'Huillier,
  and Lewenstein}}]{p_antoine_96}
\bibinfo{author}{\bibfnamefont{P.}~\bibnamefont{Antoine}},
  \bibinfo{author}{\bibfnamefont{A.}~\bibnamefont{L'Huillier}},
  \bibnamefont{and}
  \bibinfo{author}{\bibfnamefont{M.}~\bibnamefont{Lewenstein}},
  \bibinfo{journal}{Phys. Rev. Lett.} \textbf{\bibinfo{volume}{77}},
  \bibinfo{pages}{1234} (\bibinfo{year}{1996}).

\bibitem[{\citenamefont{Paul et~al.}(2001)\citenamefont{Paul, Toma, Breger,
  Mullot, Augé, Balcou, Muller, and Agostini}}]{p_paul_01}
\bibinfo{author}{\bibfnamefont{P.~M.} \bibnamefont{Paul}},
  \bibinfo{author}{\bibfnamefont{E.~S.} \bibnamefont{Toma}},
  \bibinfo{author}{\bibfnamefont{P.}~\bibnamefont{Breger}},
  \bibinfo{author}{\bibfnamefont{G.}~\bibnamefont{Mullot}},
  \bibinfo{author}{\bibfnamefont{F.}~\bibnamefont{Augé}},
  \bibinfo{author}{\bibfnamefont{P.}~\bibnamefont{Balcou}},
  \bibinfo{author}{\bibfnamefont{H.~G.} \bibnamefont{Muller}},
  \bibnamefont{and} \bibinfo{author}{\bibfnamefont{P.}~\bibnamefont{Agostini}},
  \bibinfo{journal}{Science} \textbf{\bibinfo{volume}{292}},
  \bibinfo{pages}{1689} (\bibinfo{year}{2001}).

\bibitem[{\citenamefont{Drescher et~al.}(2001)\citenamefont{Drescher,
  Hentschel, Kienberger, Tempea, Spielmann, Reider, Corkum, and
  Krausz}}]{m_drescher_01}
\bibinfo{author}{\bibfnamefont{M.}~\bibnamefont{Drescher}},
  \bibinfo{author}{\bibfnamefont{M.}~\bibnamefont{Hentschel}},
  \bibinfo{author}{\bibfnamefont{R.}~\bibnamefont{Kienberger}},
  \bibinfo{author}{\bibfnamefont{G.}~\bibnamefont{Tempea}},
  \bibinfo{author}{\bibfnamefont{C.}~\bibnamefont{Spielmann}},
  \bibinfo{author}{\bibfnamefont{G.~A.} \bibnamefont{Reider}},
  \bibinfo{author}{\bibfnamefont{P.~B.} \bibnamefont{Corkum}},
  \bibnamefont{and} \bibinfo{author}{\bibfnamefont{F.}~\bibnamefont{Krausz}},
  \bibinfo{journal}{Science} \textbf{\bibinfo{volume}{291}},
  \bibinfo{pages}{1923} (\bibinfo{year}{2001}).

\bibitem[{\citenamefont{Krause et~al.}(1992)\citenamefont{Krause, Schafer, and
  Kulander}}]{j_krause_92}
\bibinfo{author}{\bibfnamefont{J.~L.} \bibnamefont{Krause}},
  \bibinfo{author}{\bibfnamefont{K.~J.} \bibnamefont{Schafer}},
  \bibnamefont{and} \bibinfo{author}{\bibfnamefont{K.~C.}
  \bibnamefont{Kulander}}, \bibinfo{journal}{Phys. Rev. Lett.}
  \textbf{\bibinfo{volume}{68}}, \bibinfo{pages}{3535} (\bibinfo{year}{1992}).

\bibitem[{\citenamefont{{Lewenstein} et~al.}(1994)\citenamefont{{Lewenstein},
  {Balcou}, {Ivanov}, {L'huillier}, and {Corkum}}}]{m_lewenstein_94}
\bibinfo{author}{\bibfnamefont{M.}~\bibnamefont{{Lewenstein}}},
  \bibinfo{author}{\bibfnamefont{P.}~\bibnamefont{{Balcou}}},
  \bibinfo{author}{\bibfnamefont{M.~Y.} \bibnamefont{{Ivanov}}},
  \bibinfo{author}{\bibfnamefont{A.}~\bibnamefont{{L'huillier}}},
  \bibnamefont{and} \bibinfo{author}{\bibfnamefont{P.~B.}
  \bibnamefont{{Corkum}}}, \bibinfo{journal}{\pra}
  \textbf{\bibinfo{volume}{49}}, \bibinfo{pages}{2117} (\bibinfo{year}{1994}).

\bibitem[{\citenamefont{{Ghimire} et~al.}(2011)\citenamefont{{Ghimire},
  {Dichiara}, {Sistrunk}, {Agostini}, {Dimauro}, and {Reis}}}]{s_ghimire_11}
\bibinfo{author}{\bibfnamefont{S.}~\bibnamefont{{Ghimire}}},
  \bibinfo{author}{\bibfnamefont{A.~D.} \bibnamefont{{Dichiara}}},
  \bibinfo{author}{\bibfnamefont{E.}~\bibnamefont{{Sistrunk}}},
  \bibinfo{author}{\bibfnamefont{P.}~\bibnamefont{{Agostini}}},
  \bibinfo{author}{\bibfnamefont{L.~F.} \bibnamefont{{Dimauro}}},
  \bibnamefont{and} \bibinfo{author}{\bibfnamefont{D.~A.}
  \bibnamefont{{Reis}}}, \bibinfo{journal}{Nat. Phys.}
  \textbf{\bibinfo{volume}{7}}, \bibinfo{pages}{138} (\bibinfo{year}{2011}).

\bibitem[{\citenamefont{Nenciu}(1991)}]{g_nenciu_91}
\bibinfo{author}{\bibfnamefont{G.}~\bibnamefont{Nenciu}},
  \bibinfo{journal}{Rev. Mod. Phys} \textbf{\bibinfo{volume}{63}},
  \bibinfo{pages}{91} (\bibinfo{year}{1991}).

\bibitem[{\citenamefont{Davies and Wilkins}(1988)}]{j_davies_88}
\bibinfo{author}{\bibfnamefont{J.~H.} \bibnamefont{Davies}} \bibnamefont{and}
  \bibinfo{author}{\bibfnamefont{J.~W.} \bibnamefont{Wilkins}},
  \bibinfo{journal}{\prb} \textbf{\bibinfo{volume}{38}}, \bibinfo{pages}{1667}
  (\bibinfo{year}{1988}).

\bibitem[{\citenamefont{Mahan}(1990)}]{g_mahan}
\bibinfo{author}{\bibfnamefont{G.}~\bibnamefont{Mahan}},
  \emph{\bibinfo{title}{Many-Particle Physics}} (\bibinfo{publisher}{Plenum
  Press, NY}, \bibinfo{year}{1990}).

\bibitem[{\citenamefont{Falicov and Kimball}(1969)}]{l_falicov_69}
\bibinfo{author}{\bibfnamefont{L.~M.} \bibnamefont{Falicov}} \bibnamefont{and}
  \bibinfo{author}{\bibfnamefont{J.~C.} \bibnamefont{Kimball}},
  \bibinfo{journal}{\prl} \textbf{\bibinfo{volume}{22}}, \bibinfo{pages}{997}
  (\bibinfo{year}{1969}).

\bibitem[{\citenamefont{{Peierls}}(1933)}]{r_peierls_33}
\bibinfo{author}{\bibfnamefont{R.}~\bibnamefont{{Peierls}}},
  \bibinfo{journal}{Zeitschrift fur Physik} \textbf{\bibinfo{volume}{80}},
  \bibinfo{pages}{763} (\bibinfo{year}{1933}).

\bibitem[{\citenamefont{{Jauho} and {Wilkins}}(1984)}]{a_jauho_84}
\bibinfo{author}{\bibfnamefont{A.~P.} \bibnamefont{{Jauho}}} \bibnamefont{and}
  \bibinfo{author}{\bibfnamefont{J.~W.} \bibnamefont{{Wilkins}}},
  \bibinfo{journal}{\prb} \textbf{\bibinfo{volume}{29}}, \bibinfo{pages}{1919}
  (\bibinfo{year}{1984}).

\bibitem[{\citenamefont{{Holstein}}(1959)}]{t_holstein_59}
\bibinfo{author}{\bibfnamefont{T.}~\bibnamefont{{Holstein}}},
  \bibinfo{journal}{Annals of Physics} \textbf{\bibinfo{volume}{8}},
  \bibinfo{pages}{325} (\bibinfo{year}{1959}).

\bibitem[{\citenamefont{Akima}(1970)}]{h_akima_70}
\bibinfo{author}{\bibfnamefont{H.}~\bibnamefont{Akima}}, \bibinfo{journal}{J.
  ACM} \textbf{\bibinfo{volume}{17}}, \bibinfo{pages}{589}
  (\bibinfo{year}{1970}).

\bibitem[{\citenamefont{Press et~al.}(2007)\citenamefont{Press, Teukolsky,
  Vetterling, and Flannery}}]{numerical_recipes}
\bibinfo{author}{\bibfnamefont{W.~H.} \bibnamefont{Press}},
  \bibinfo{author}{\bibfnamefont{S.~A.} \bibnamefont{Teukolsky}},
  \bibinfo{author}{\bibfnamefont{W.~T.} \bibnamefont{Vetterling}},
  \bibnamefont{and} \bibinfo{author}{\bibfnamefont{B.~P.}
  \bibnamefont{Flannery}}, \emph{\bibinfo{title}{Numerical Recipes 3rd Edition:
  The Art of Scientific Computing}} (\bibinfo{publisher}{Cambridge University
  Press}, \bibinfo{address}{New York, NY, USA}, \bibinfo{year}{2007}),
  \bibinfo{edition}{3rd} ed., ISBN \bibinfo{isbn}{0521880688, 9780521880688}.

\bibitem[{\citenamefont{Kuzmin and Haviland}(1991)}]{l_kuzmin_91}
\bibinfo{author}{\bibfnamefont{L.}~\bibnamefont{Kuzmin}} \bibnamefont{and}
  \bibinfo{author}{\bibfnamefont{D.}~\bibnamefont{Haviland}},
  \bibinfo{journal}{\prl} \textbf{\bibinfo{volume}{67}}, \bibinfo{pages}{2890}
  (\bibinfo{year}{1991}).

\bibitem[{\citenamefont{Waschke et~al.}(1993)\citenamefont{Waschke, Roskos,
  Schwedler, Leo, Kurz, and K{\"o}hler}}]{c_waschke_93}
\bibinfo{author}{\bibfnamefont{C.}~\bibnamefont{Waschke}},
  \bibinfo{author}{\bibfnamefont{H.}~\bibnamefont{Roskos}},
  \bibinfo{author}{\bibfnamefont{R.}~\bibnamefont{Schwedler}},
  \bibinfo{author}{\bibfnamefont{K.}~\bibnamefont{Leo}},
  \bibinfo{author}{\bibfnamefont{H.}~\bibnamefont{Kurz}}, \bibnamefont{and}
  \bibinfo{author}{\bibfnamefont{K.}~\bibnamefont{K{\"o}hler}},
  \bibinfo{journal}{\prl} \textbf{\bibinfo{volume}{70}}, \bibinfo{pages}{3319}
  (\bibinfo{year}{1993}).

\bibitem[{\citenamefont{Lyssenko et~al.}(1997)\citenamefont{Lyssenko, Valu{\v
  s}is, L{\"o}ser, and Hasche}}]{v_lyssenko_97}
\bibinfo{author}{\bibfnamefont{V.}~\bibnamefont{Lyssenko}},
  \bibinfo{author}{\bibfnamefont{G.}~\bibnamefont{Valu{\v s}is}},
  \bibinfo{author}{\bibfnamefont{F.}~\bibnamefont{L{\"o}ser}},
  \bibnamefont{and} \bibinfo{author}{\bibfnamefont{T.}~\bibnamefont{Hasche}},
  \bibinfo{journal}{\prl} \textbf{\bibinfo{volume}{79}}, \bibinfo{pages}{301}
  (\bibinfo{year}{1997}).

\bibitem[{\citenamefont{Ben~Dahan et~al.}(1996)\citenamefont{Ben~Dahan, Peik,
  Reichel, Castin, and Salomon}}]{m_bendahan_96}
\bibinfo{author}{\bibfnamefont{M.}~\bibnamefont{Ben~Dahan}},
  \bibinfo{author}{\bibfnamefont{E.}~\bibnamefont{Peik}},
  \bibinfo{author}{\bibfnamefont{J.}~\bibnamefont{Reichel}},
  \bibinfo{author}{\bibfnamefont{Y.}~\bibnamefont{Castin}}, \bibnamefont{and}
  \bibinfo{author}{\bibfnamefont{C.}~\bibnamefont{Salomon}},
  \bibinfo{journal}{\prl} \textbf{\bibinfo{volume}{76}}, \bibinfo{pages}{4508}
  (\bibinfo{year}{1996}).

\bibitem[{\citenamefont{{Ghimire} et~al.}(2012)\citenamefont{{Ghimire},
  {DiChiara}, {Sistrunk}, {Ndabashimiye}, {Szafruga}, {Mohammad}, {Agostini},
  {DiMauro}, and {Reis}}}]{s_ghimire_12}
\bibinfo{author}{\bibfnamefont{S.}~\bibnamefont{{Ghimire}}},
  \bibinfo{author}{\bibfnamefont{A.~D.} \bibnamefont{{DiChiara}}},
  \bibinfo{author}{\bibfnamefont{E.}~\bibnamefont{{Sistrunk}}},
  \bibinfo{author}{\bibfnamefont{G.}~\bibnamefont{{Ndabashimiye}}},
  \bibinfo{author}{\bibfnamefont{U.~B.} \bibnamefont{{Szafruga}}},
  \bibinfo{author}{\bibfnamefont{A.}~\bibnamefont{{Mohammad}}},
  \bibinfo{author}{\bibfnamefont{P.}~\bibnamefont{{Agostini}}},
  \bibinfo{author}{\bibfnamefont{L.~F.} \bibnamefont{{DiMauro}}},
  \bibnamefont{and} \bibinfo{author}{\bibfnamefont{D.~A.}
  \bibnamefont{{Reis}}}, \bibinfo{journal}{\pra} \textbf{\bibinfo{volume}{85}},
  \bibinfo{eid}{043836} (\bibinfo{year}{2012}).

\bibitem[{\citenamefont{Zhukov and Chulkov}(2010)}]{v_zhukov_10}
\bibinfo{author}{\bibfnamefont{V.}~\bibnamefont{Zhukov}} \bibnamefont{and}
  \bibinfo{author}{\bibfnamefont{E.}~\bibnamefont{Chulkov}},
  \bibinfo{journal}{J. Phys. Cond. Matter} \textbf{\bibinfo{volume}{22}},
  \bibinfo{pages}{435802} (\bibinfo{year}{2010}).

\bibitem[{\citenamefont{Lindenberg}(2011)}]{a_lindenberg_priv_11}
\bibinfo{author}{\bibfnamefont{A.}~\bibnamefont{Lindenberg}},
  \bibinfo{howpublished}{private communication} (\bibinfo{year}{2011}).

\bibitem[{\citenamefont{Gr\"ochenig}(2001)}]{gabor}
\bibinfo{author}{\bibfnamefont{K.}~\bibnamefont{Gr\"ochenig}},
  \emph{\bibinfo{title}{Foundations of Time-Frequency Analysis}}
  (\bibinfo{publisher}{Birkh\"auser}, \bibinfo{year}{2001}).

\bibitem[{\citenamefont{{Vogl}}(1983)}]{p_vogl_83}
\bibinfo{author}{\bibfnamefont{P.}~\bibnamefont{{Vogl}}},
  \bibinfo{journal}{Journal of Physics and Chemistry of Solids}
  \textbf{\bibinfo{volume}{44}}, \bibinfo{pages}{365} (\bibinfo{year}{1983}).

\bibitem[{\citenamefont{{Freericks} et~al.}(2012)\citenamefont{{Freericks},
  {Liu}, {Kemper}, and {Devereaux}}}]{j_freericks_12}
\bibinfo{author}{\bibfnamefont{J.~K.} \bibnamefont{{Freericks}}},
  \bibinfo{author}{\bibfnamefont{A.~Y.} \bibnamefont{{Liu}}},
  \bibinfo{author}{\bibfnamefont{A.~F.} \bibnamefont{{Kemper}}},
  \bibnamefont{and} \bibinfo{author}{\bibfnamefont{T.~P.}
  \bibnamefont{{Devereaux}}}, \bibinfo{journal}{Physica Scripta Volume T}
  \textbf{\bibinfo{volume}{151}}, \bibinfo{pages}{014062}
  (\bibinfo{year}{2012}), \eprint{1212.1610}.

\end{thebibliography}

\end{document}